\documentclass[prb,amsmath,amssymb,twocolumn,longbibliography]{revtex4-2}
\usepackage{graphicx}
\usepackage{dcolumn}
\usepackage{bm}
\usepackage{placeins}
\usepackage{rotating}
\usepackage{multirow}
\usepackage[dvipsnames,usenames]{xcolor}
\usepackage[english]{babel}
\usepackage{amssymb}
\usepackage{amsmath}
\usepackage[utf8]{inputenc}
\usepackage[normalem]{ulem}
\usepackage{float}
\usepackage{array, booktabs,makecell}
\newcommand{\be}{\begin{equation}}
\newcommand{\ee}{\end{equation}}
\newcommand{\bea}{\begin{eqnarray}}
\newcommand{\eea}{\end{eqnarray}}

\newcolumntype{M}[1]{>{\centering\arraybackslash}m{#1}}
\newcolumntype{N}{@{}m{0pt}@{}}

\def\a{\alpha}

\def\G{\Gamma}
\def\d{\delta}

\def\l{\lambda}

\def\w{\omega}

\begin{document}

\title{Atomic forces from correlation energy functionals based on the adiabatic-\\connection fluctuation-dissipation theorem}
\author{Damian Contant}
\affiliation{Sorbonne Universit\'e, MNHN, UMR CNRS 7590, IMPMC, 4 place Jussieu, 75005 Paris, France}
\author{Maria Hellgren}
\affiliation{Sorbonne Universit\'e, MNHN, UMR CNRS 7590, IMPMC, 4 place Jussieu, 75005 Paris, France}
\date{\today}
\begin{abstract}
We extend the capabilities of correlation energy functionals based on the adiabatic-connection fluctuation-dissipation theorem by implementing the analytical atomic forces within the random phase approximation (RPA), in the context of plane waves and pseudopotentials. Forces are calculated at self-consistency through the optimized effective potential method and the Hellmann-Feynman theorem. In addition, non-self-consistent RPA forces, starting from the PBE generalized gradient approximation, are evaluated using density functional perturbation theory. In both cases, we find forces of excellent numerical quality. Furthermore, for most molecules and solids studied, self-consistency is found to have a negligible impact on the computed geometries and vibrational frequencies. The RPA is shown to systematically improve over PBE and, by including the exact-exchange kernel within RPA \!+\! exchange (RPAx), through finite-difference total energy calculations, we obtain an accuracy comparable to advanced wavefunction methods. Finally, we estimate the anharmonic shift and provide accurate theoretical references based on RPA and RPAx for the zone-center optical phonon of diamond, silicon, and germanium.
\end{abstract}
\maketitle

\section{Introduction}
The Born-Oppenheimer (BO) approximation \cite{born1927ZurQuantentheorieMolekeln} combined with Kohn-Sham (KS) density functional theory (DFT) \cite{hohenberg1964InhomogeneousElectronGas,kohn1965SelfConsistentEquationsIncluding} provides a comprehensive theoretical framework for first principles simulations of molecules and materials. Effects of the electron-electron interaction are incorporated, in principle exactly, through the so-called exchange correlation (xc) energy functional \cite{perdew_jacobsladder2001}. In most applications, the xc functional is approximated in the local density approximation (LDA) or the semi-local generalized gradient approximation (GGA). These types of approximations are computationally cheap but suffer from well-known drawbacks due to electronic self-interaction and missing van der Waals forces \cite{RappoportBurke:09}.

In this regard, functionals based on the adiabatic-connection fluctuation-dissipation theorem (ACFDT) \cite{langreth1975ExchangecorrelationEnergyMetallic,langreth1977ExchangecorrelationEnergyMetallic}, such as the random phase approximation (RPA) \cite{ren2012RandomphaseApproximationIts,eshuis2012ElectronCorrelationMethods} and the RPA + exchange (RPAx) \cite{hellgren2010,Hesselmann10022010,colonna2014CorrelationEnergyExactexchange,bleiziffer2015,hellgren2018RandomPhaseApproximation}, have shown to offer a significant and reliable improvement, with results often of similar quality to advanced wavefunction methods. The interlayer binding energy of layered materials, such as graphite and layered boron nitride, is well-reproduced \cite{PhysRevLett.105.196401,PhysRevLett.96.136404}, and small energy differences between polymorphic solids are captured with high accuracy \cite{Bing2012,hellgren2021RandomPhaseApproximation}. Recent calculations on high-pressure molecular solid hydrogen phases showed excellent agreement with diffusion Monte Carlo \cite{hellgren2022HighpressureIIIIIPhase}. 
  
The performance with respect to structural and vibrational properties is, so far, less explored. A common practice is to employ a more advanced approximation to the electronic energy and combine it with 
optimized structures and vibrational properties calculated with a GGA functional, usually within the harmonic approximation. Although this often leads to useful results, semi-local approximations tend to overestimate bond distances and underestimate vibrational frequencies, problems which become increasingly more severe close to a second-order structural phase transition \cite{lazzeri2008,barborini2015,hellgrentise2_2017,hellgren_CNT2018,hellgren2022HighpressureIIIIIPhase}. It is, therefore, motivated to extend the capabilities of functionals based on the ACFDT for describing the coupling between lattice and electronic degrees of freedom.
Staying within the harmonic approximation, it is then necessary to evaluate the analytical forces, i.e., the first-order derivatives of the BO potential energy surface 
with respect to atomic displacements. 

The analytical forces in the RPA have been considered in a few works. In Refs. \cite{rekkedal2013CommunicationAnalyticGradients,burow2014AnalyticalFirstOrderMolecular,mussard2014AnalyticalEnergyGradients,tahir2022LocalizedResolutionIdentity,stein2024MassivelyParallelImplementation,bates2025FrozenCoreAnalyticalGradients,tahir2025AnalyticalGradientsRandomPhase}, the analytical non-self-consistent RPA forces (i.e. the analytical gradient of the RPA energy evaluated with orbitals and eigenvalues from a semi-local approximation) were implemented in the context of Gaussian basis sets, and tested on a variety of molecular systems. Recently, Thierbach et al.  \cite{thierbach2020AnalyticEnergyGradientsa,trushin2021NumericallyStableOptimized} used Gaussian basis sets to evaluate RPA forces at self-consistency, solving the optimized effective potential (OEP) equation \cite{sharp1953VariationalApproachUnipotential,talman1976OptimizedEffectiveAtomic,casida1995GeneralizationOptimizedeffectivepotentialModel,hellgren2007CorrelationPotentialDensity,nguyen2014InitioSelfconsistentTotalenergy,Riemelmoser2021,pitts2025SelfconsistentRandomPhase,Trushin2025}. 
While Gaussian functions are efficient for describing finite-size systems such as molecules, the use of plane waves and pseudopotentials is often more convenient in solids. 
To date, only one work has reported non-self-consistent RPA forces with plane waves in the projector-augmented-wave framework \cite{ramberger2017AnalyticInteratomicForces}. 

In this work,  we derive analytical expressions for the self-consistent and non-self-consistent evaluation of RPA forces, in the presence of nonlocal external potentials, such as those arising from norm-conserving pseudopotentials \cite{contant2024OptimizedEffectivePotential}. The force expressions are implemented in the context of plane-wave basis sets within the \texttt{ACFDT} package \cite{nguyen2009,nguyen2014InitioSelfconsistentTotalenergy,colonna2014CorrelationEnergyExactexchange,Colonna2016,hellgren2018RandomPhaseApproximation,hellgren2021RandomPhaseApproximation,pitts2025SelfconsistentRandomPhase} of the \texttt{Quantum} \texttt{ESPRESSO} (\texttt{QE}) distribution \cite{giannozzi2009QUANTUMESPRESSOModular,giannozzi2017AdvancedCapabilitiesMaterials}, which exploits an eigenvalue decomposition of the linear density response function for evaluating the RPA energy and potential. 
We demonstrate that our implementation achieves a high numerical quality, enabling the exploitation of RPA forces in a wide range of applications. We also determine forces within RPAx using finite-difference total energy calculations. The accuracy of the different methods is assessed by comparing structural and vibrational properties of molecules and solids to published results obtained with coupled-cluster single, double, and perturbative triple excitations (CCSD(T)), Diffusion Monte Carlo (DMC), and experiments.   

The paper is organized as follows. In Sec. II, we recall the ACFDT total energy expression. In Sec. III, we derive the RPA atomic forces in detail and discuss their implementation. In Sec. IV, we demonstrate the numerical accuracy of the calculated forces. In Sec. V, we present the performance of the RPA and RPAx methods in predicting structural and vibrational properties of a set of molecules (H$_2$, LiH, HF, N$_2$, H$_2$O, NH$_3$, and CH$_4$) and solids (C, Si, and Ge). Finally, we present our conclusions in Sec. VI.

\section{ACFDT formalism}
Within the Born-Oppenheimer (BO) approximation, the potential energy surface is given by the sum of the following contributions: the internuclear interaction energy, $E_{\text{NN}}$, the electronic kinetic energy, $T_{e}$, the electron-nuclei (external) interaction energy, $E_{\text{ext}}$, and the electron-electron interaction energy, $W$. Within the KS-DFT framework, the sum of $W$ and $T_{e}$ is re-expressed as a sum of the kinetic energy of independent KS electrons, $T^{\text{KS}}_{e}$, the Hartree energy, $E_{\text{H}}$, the exchange energy, $E_{\text{x}}$, and the correlation energy, $E_{\text{c}}$. Typically, as within the LDA and GGA, both $E_{\text{x}}$ and $E_{\text{c}}$ are approximated. 

Methods such as the RPA and RPAx have an exact description of the exchange energy, while the correlation energy is approximated starting from the formally exact ACFDT formula \cite{langreth1975ExchangecorrelationEnergyMetallic,langreth1977ExchangecorrelationEnergyMetallic},
\begin{equation}
\begin{aligned}
\displaystyle E_{\text{c}}^{\text{ACFDT}} = \int \limits_{0}^{1} d\lambda \int \limits_0^{\infty}  \frac{du}{2\pi} \, \text{Tr} \bigl \{& v  \bigl  [ \chi_{\lambda} (iu)  - \chi_{\text{0}}(iu) \bigr ] \bigr \}.
\label{eq:acfdt_correlation}
\end{aligned}
\end{equation}
Here, $v$ is the bare Coulomb interaction and $\lambda$ is a parameter connecting the fictitious KS system of independent particles ($\lambda \!= \!0$) to the physical system ($\lambda \!= \!1$), while keeping the charge density fixed at $\l=1$. 

The key ingredient in the ACFDT formula is the linear density response function, evaluated for the $\l$-dependent system and for the KS system. The latter is defined by the one-particle KS potential
\begin{equation}
v_{\text{eff}}(\mathbf{r})=v_{\text{ext}}(\mathbf{r})+v_{\text{H}}(\mathbf{r})+v_{\text{xc}}(\mathbf{r}). 
\label{eq:acfdt_kspotential}
\end{equation}
The two response functions, $\chi_{\lambda}$ and $\chi_{\text{0}}$, are connected through the Dyson equation of time-dependent density functional theory \cite{tddftbook},
\begin{equation}
\displaystyle \chi_{\lambda} = \chi_{\text{0}} + \chi_{\text{0}} \Bigl [ \lambda v + f_{\text{xc}}^{\lambda} \Bigr ] \chi_{\lambda}.
\label{eq:acfdt_response}
\end{equation}
Here, $f^{\lambda}_{\text{xc}}$ corresponds to the xc kernel, defined as the functional derivative of the time-dependent xc potential, $v_{\text{xc}}^{\lambda}$, with respect to the electronic density, $n$ \cite{gross1985}. 
Neglecting the exchange and correlation parts of the kernel altogether results in the RPA, in which $\chi_\l$ is approximated by the time-dependent Hartree approximation. Including exact exchange in the xc kernel ($f^{\l}_{\rm xc}\approx f^{\l}_{\rm x}$) defines the RPAx approximation \cite{hellgren2010,Hesselmann10022010,colonna2014CorrelationEnergyExactexchange,bleiziffer2015,hellgren2018RandomPhaseApproximation}. The exact-exchange response kernel completes the second-order terms in the correlation energy, and generates higher-order "exchange" contributions through Eq.~(\ref{eq:acfdt_response})  \cite{hellgren2018RandomPhaseApproximation}.

The Hartree and exact-exchange kernels are both linear in $\l$ ($f^\l_{\text{x}}=\l f_{\text{x}}$). It is, therefore, possible to integrate over $\l$ in Eq.~(\ref{eq:acfdt_correlation}) analytically. Combining Eqs.~(\ref{eq:acfdt_correlation}) and (\ref{eq:acfdt_response}) gives the following expression for the correlation energy in RPAx \cite{hellgren2010} (or RPA, by setting $f_{\text{x}}=0$)
 \begin{equation}
 \begin{aligned}
\displaystyle E_{\text{c}}^{\text{RPAx}} =& \int \limits_{0}^{\infty} \frac{du}{2\pi} \, \text{Tr} \left\{ v \chi_{\text{0}}(iu)+\frac{v}{v+f_{\rm x}(iu)}\right.\\
&\,\,\,\,\,\,\,\,\,\,\,\,\,\,\,\,\,\,\,\,\,\,\,\,\,\,\,\left.\times\ln\left[1- [v+f_{\rm x}(iu)] \chi_{\text{0}}(iu)\right]  \right\}.
\label{eq:rpa_acfdt_correlation_practice}
\end{aligned}
\end{equation}
Note here that $f_{\text{x}}$ is frequency-dependent \cite{hellgren2023AdiabaticKernel}. 
In Refs. \cite{colonna2014CorrelationEnergyExactexchange,hellgren2018RandomPhaseApproximation}, the use of alternative partial resummations of the RPAx was introduced. Using the same $f_\text{x}$-kernel, we can define RPAx(1) from
\begin{equation}
\displaystyle \chi_{\lambda} = P^{(1)}_{\l} + \l P^{(1)}_{\l}  v   \chi_{\lambda},\,\,\,\,\,\,P^{(1)}_{\l}=\chi_0 + \l\chi_0 f_{\text{x}}\chi_0.
\label{eq:rpax1}
\end{equation}
This resummation can be identified with an approximation within many-body perturbation theory that includes exchange up to first order in the vertex function. 
A second variant, AC-SOSEX (adiabatic-connection-Second Order Screened EXchange), can be defined from 
\begin{equation}
\displaystyle \chi_{\lambda} = P^{(1)}_{\l} + \l  \chi_0  v   \chi_{\lambda},
\label{eq:sosex}
\end{equation}
which corresponds to a truncation of the RPAx(1) series that omits exchange in the screened interaction. Once the $f_\text{x}$-kernel is calculated, both variants are straightforwardly obtained. They usually give very similar results to RPAx but, in certain critical cases, their use is preferred. In addition to RPAx(1) and AC-SOSEX, other resummations, such as the AXK (approximate exchange kernel) have been found to be accurate \cite{bates2013CommunicationRandomPhase,chen2018PerformanceScopePerturbative}.

The response functions involve summations over excitation functions, i.e., products of occupied and unoccupied KS states. In Ref.~\cite{nguyen2009}, it was shown that the use of unoccupied states can be completely avoided for the implementation of the RPA correlation energy by solving the following eigenvalue problem
\begin{equation}
\displaystyle v \chi_{\text{0}}(iu) | \Delta V_{\alpha} \rangle = a_{\alpha} | \Delta V_{\alpha} \rangle.
\label{eq:energy_rpa_eigenvalue_problem}
\end{equation}
Solving Eq.~(\ref{eq:energy_rpa_eigenvalue_problem}) for a given imaginary frequency returns a set of eigenpotentials $\{V_{\alpha}\}$ and corresponding eigenvalues $\{a_{\alpha}\}$. The RPA correlation energy can then be expressed as a sum involving the eigenvalues only,
\begin{equation}
\displaystyle E_{\text{c}}^{\text{RPA}} = \sum \limits_{\alpha}^{N_{\text{eig}}} \int \limits_{0}^{\infty} \frac{du}{2\pi} \ln\left[1 - a_{\alpha}(iu)\right] + a_{\alpha}(iu).
\label{eq:rpa_acfdt_correlation_eigenvalues}
\end{equation}
The eigenvalue problem in Eq.~(\ref{eq:energy_rpa_eigenvalue_problem}) can be solved iteratively by treating the eigenpotentials as first-order perturbations acting on the electronic ground state \cite{nguyen2009}, a task that can be effectively handled using the formalism of density functional perturbation theory (DFPT) \cite{sternheimer1954ElectronicPolarizabilitiesIons,baroni2001PhononsRelatedCrystal}. A similar eigenvalue problem involving the $f_{\text{x}}$-kernel can be formulated for RPAx and its variants (see Refs. \cite{colonna2014CorrelationEnergyExactexchange, hellgren2021RandomPhaseApproximation} for more details).

Determining the RPA and RPAx correlation energy requires a starting point. Such non-self-consistent calculations always introduce a bias, which can only be completely suppressed by identifying the optimal KS starting point. This is accomplished through the OEP approach  \cite{sharp1953VariationalApproachUnipotential,talman1976OptimizedEffectiveAtomic,casida1995GeneralizationOptimizedeffectivepotentialModel,hellgren2007CorrelationPotentialDensity,nguyen2014InitioSelfconsistentTotalenergy,Riemelmoser2021,pitts2025SelfconsistentRandomPhase,Trushin2025}, which allows orbital-dependent functionals to work within the standard Kohn-Sham scheme of DFT \cite{kummel2008OrbitaldependentDensityFunctionals}. It applies not only to RPA but also to any method that includes exact exchange, such as hybrid functionals. For a given xc functional, the OEP equation reads
\begin{equation}
\displaystyle \frac{\delta E_{\text{xc}}}{\delta v_{\text{eff}}(\mathbf{r})} = \int \frac{\delta n(\mathbf{r}')}{\delta v_{\text{eff}}(\mathbf{r})} \, v_{\text{xc}}(\mathbf{r}') \, d\mathbf{r}'.
\label{eq:oep_equation_xc}
\end{equation}
Both the left- and right-hand sides of Eq.~(\ref{eq:oep_equation_xc}) can be written in terms of density responses. This feature was exploited in Ref. \cite{nguyen2014InitioSelfconsistentTotalenergy} to solve the OEP equation, combining DFPT with an iterative algorithm for extracting the xc potential. The RPA correlation term contributing to the left-hand side of Eq.~(\ref{eq:oep_equation_xc}) can be evaluated as 
\begin{equation}
\displaystyle \frac{\delta E_{\text{c}}^{\text{RPA}}}{\delta v_{\text{eff}}(\mathbf{r})} =  \int \limits_{0}^{\infty} \frac{du}{2\pi} \, \sum_{\alpha}^{N_{\alpha}} \, \frac{- a_{\alpha}(iu) }{1 - a_{\alpha}(iu)} \frac{\delta a_{\alpha}(iu)}{\delta v_{\text{eff}}(\mathbf{r})}.
\label{eq:energy_correlation_rpa_develop}
\end{equation}
The variation of each eigenvalue with respect to $v_{\text{eff}}$ 
can be further split into different terms, which can all be evaluated using DFPT \cite{nguyen2014InitioSelfconsistentTotalenergy,pitts2025SelfconsistentRandomPhase}.

Following the eigenvalue problem in Eq.~(\ref{eq:energy_rpa_eigenvalue_problem}), the total energy in RPA/RPAx and the RPA potential have all been implemented within the \texttt{ACFDT} code \cite{nguyen2009,nguyen2014InitioSelfconsistentTotalenergy,colonna2014CorrelationEnergyExactexchange,Colonna2016,hellgren2018RandomPhaseApproximation,hellgren2021RandomPhaseApproximation,pitts2025SelfconsistentRandomPhase} of the  \texttt{QE} distribution \cite{giannozzi2009QUANTUMESPRESSOModular,giannozzi2017AdvancedCapabilitiesMaterials}. Building on these capabilities, we have, in this work, implemented the force expression, as described in the next section.
\section{Atomic forces in RPA}
Having established a way to calculate the total energy, we can now look at its first-order derivatives with respect to the atomic positions. This yields a set of forces, which are important for studying the stability and excitations of a lattice or molecular structure. The force, $\mathbf{F}_{I}$, exerted on a given atom, $I$, corresponds to the derivative of the BO potential energy surface with respect to the atomic position, $\mathbf{R}_{I}$. Thus, from Sec. II, we have
\begin{equation}
\displaystyle \mathbf{F}_{I} = - \frac{\partial E_{\text{NN}}}{\partial \mathbf{R}_{I}} - \frac{\partial T^{\text{KS}}_{e}}{\partial \mathbf{R}_{I}} - \frac{\partial E_{\text{ext}}}{\partial \mathbf{R}_{I}} - \frac{\partial E_{\text{H}}}{\partial \mathbf{R}_{I}} - \frac{\partial E_{\text{xc}}}{\partial \mathbf{R}_{I}}.
\label{eq:force_total}
\end{equation}
By evaluating these energy derivatives through an electronic KS density, generated by $v_{\text{eff}}(\mathbf{r})$ (see Eq.~(\ref{eq:acfdt_kspotential})), we obtain the following general expression for the force exerted on atom $I$
\begin{equation}
\begin{aligned}
\displaystyle \mathbf{F}_{I} = & - \frac{\partial}{\partial \mathbf{R}_{I}} \sum \limits_{A \neq I} \frac{Z_{A} Z_{I}}{|\mathbf{R}_{A} - \mathbf{R}_{I}|} - \int \! n(\mathbf{r}) \frac{\partial v_{\text{ext}}(\mathbf{r})}{\partial \mathbf{R}_{I}} \, d\mathbf{r} \\
& + \int\! v_{\text{xc}}(\mathbf{r}) \frac{\partial n(\mathbf{r})}{\partial \mathbf{R}_{I}} \, d\mathbf{r} - \frac{\partial E_{\text{xc}}}{\partial \mathbf{R}_{I}},
\label{eq:force_total_developed}
\end{aligned}
\end{equation}
where $Z_{I}$ corresponds to the nuclear charge of atom $I$, and 
\begin{equation}
\begin{aligned}
\displaystyle v_{\text{ext}}(\mathbf{r}) =  \sum \limits_{A} \frac{Z_{A}}{|\mathbf{R}_{A} - \mathbf{r}|}
\label{eq:force_external}
\end{aligned}
\end{equation}
is the external (nuclear) potential. The first term in Eq.~(\ref{eq:force_total_developed}) corresponds to the Ewald contribution, which is linked to the internuclear repulsion \cite{ewald1921BerechnungOptischerUnd}. The second term describes the interaction between the nuclei and the electronic density. The third and fourth terms are the only ones that depend directly on the xc functional. In the next two subsections, we will adapt Eq.~(\ref{eq:force_total_developed}) to RPA at self-consistency and to non-self-consistent RPA, starting from a local or semi-local approximation.

\subsection{Forces at self-consistency}
As mentioned previously, in order to avoid starting-point biases, a self-consistent (scf) solution should ideally be considered. Furthermore, self-consistency simplifies the expression for the force given in Eq.~(\ref{eq:force_total_developed}), thanks to the validity of the Hellmann-Feynman theorem (HFT) \cite{hellmann1937EinfuehrungQuantenchemie,feynman1939ForcesMolecules}. At self-consistency, the xc potential is equal to the functional derivative of the xc energy with respect to the charge density. By applying the chain rule on the fourth term of Eq.~(\ref{eq:force_total_developed}) and using the fact that, at self-consistency, the density and xc potential are identical in both the third and fourth term, it is easy to see that these two contributions cancel. We can thus write
\begin{equation}
\begin{aligned}
\mathbf{F}_{I}^{\text{scf}} = & \,\,\mathbf{F}_{I}^{\text{NN}} +\mathbf{F}_{I}^{\text{ext}}= \\
&- \frac{\partial}{\partial \mathbf{R}_{I}} \sum \limits_{A \neq I} \frac{Z_{A} Z_{I}}{|\mathbf{R}_{A} - \mathbf{R}_{I}|} - \int \!n_{\text{scf}}(\mathbf{r}) \frac{\partial v_{\text{ext}}(\mathbf{r})}{\partial \mathbf{R}_{I}} \, d\mathbf{r}.
\label{eq:force_hft_final}
\end{aligned}
\end{equation}
This is the textbook Hellmann-Feynman force formula, used to compute atomic forces at the LDA and GGA levels of theory.  

The HFT can also be used with orbital-dependent functionals that rely on the OEP method. This was confirmed in Refs. \cite{wu2005AnalyticEnergyGradients,thierbach2020AnalyticEnergyGradients,thierbach2020AnalyticEnergyGradientsa}, employing Gaussian basis functions. Nevertheless, very recently, it was shown that when nonlocal pseudopotentials are employed, the force expression given in Eq.~(\ref{eq:force_hft_final}) is incomplete (see Ref. \cite{contant2024OptimizedEffectivePotential}). In the pseudopotential (PP) approximation, the external potential is modified, containing both a local and, in general, a nonlocal contribution,
\begin{equation}
\begin{aligned}
v^{\text{PP}}_{\text{ext}}(\mathbf{r},\mathbf{r}')=v_{\text{L}}(\mathbf{r})\d(\mathbf{r}-\mathbf{r'})+v_{\text{NL}}(\mathbf{r},\mathbf{r}').
\label{eq:force_extpp}
\end{aligned}
\end{equation}
Within the OEP method, the charge density is optimized with respect to a strictly local potential. Thus, when the density variation comes from changes in a nonlocal potential, as is the case for the force, care must be taken when applying the chain rule in Eq.~(\ref{eq:force_total_developed}). For the third term in Eq.~(\ref{eq:force_total_developed}), we can write 
\begin{equation}
\begin{aligned}
\displaystyle \int & v_{\text{xc}}(\mathbf{r}) \frac{\partial n(\mathbf{r})}{\partial \mathbf{R}_{I}} \, d\mathbf{r} = \\
& \iiint v_{\text{xc}}(\mathbf{r}) \frac{\delta n(\mathbf{r})}{\delta v_{\text{eff}}(\mathbf{r}',\mathbf{r}'')} \frac{\partial v_{\text{eff}}(\mathbf{r}',\mathbf{r}'')}{\partial \mathbf{R}_{I}} \, d\mathbf{r} \, d\mathbf{r}' \, d\mathbf{r}''.
\label{eq:force_third_term_RPA}
\end{aligned}
\end{equation}
The effective KS potential, $v_{\text{eff}}(\mathbf{r}',\mathbf{r}'')$, is now allowed to have a (static) nonlocal component coming from Eq.~(\ref{eq:force_extpp}). Focusing on the RPA functional, the fourth term of Eq.~(\ref{eq:force_total_developed}) becomes
\begin{equation}
\begin{aligned}
\displaystyle  \frac{\partial E_{\text{xc}}^{\text{RPA}}}{\partial \mathbf{R}_{I}} = &  \iint \! \frac{\delta E_{\text{xc}}^{\text{RPA}}}{\delta  v_{\text{eff}}(\mathbf{r},\mathbf{r}')} \frac{\partial v_{\text{eff}}(\mathbf{r},\mathbf{r}')}{\partial \mathbf{R}_{I}} \, d\mathbf{r} \, d\mathbf{r}' .
\label{eq:force_fourth_term_RPA}
\end{aligned}
\end{equation}
The local components of $v_{\text{eff}}$, namely the Hartree potential, xc potential, and external local pseudopotential, will lead to terms in Eqs.~(\ref{eq:force_third_term_RPA}) and (\ref{eq:force_fourth_term_RPA}) that will cancel at self-consistency, since the OEP equation (see Eq.~(\ref{eq:oep_equation_xc})) is fulfilled. However, the terms due to the nonlocal component of the pseudopotential are not guaranteed to cancel, giving rise to a correction term of the form
\begin{equation}
\begin{aligned}
\displaystyle \Delta \mathbf{F}_{\text{xc}}^{\text{scf-RPA}} =\\
& \!\!\!\!\!\!\!\!\!\!\!\!\!\!\!\!\!\!\!\!\!\!\!\!\!\!\iiint \!v^{\text{RPA}}_{\text{xc}}(\mathbf{r}) \frac{\delta n^{\text{RPA}}(\mathbf{r})}{\delta v_{\text{eff}}(\mathbf{r}',\mathbf{r}'')} \frac{\partial v_{\text{NL}}(\mathbf{r}',\mathbf{r}'')}{\partial \mathbf{R}_{I}} \, d\mathbf{r} \, d\mathbf{r}' \, d\mathbf{r}''\\
&\!\!\!\!\!\!\!\!\!\!\!\!\!\!\!\!\!\!\!\!\!\!\!\!\!\! - \iint \! \frac{\delta E_{\text{xc}}^{\text{RPA}}}{\delta  v_{\text{eff}}(\mathbf{r},\mathbf{r}')} \frac{\partial v_{\text{NL}}(\mathbf{r},\mathbf{r}')}{\partial \mathbf{R}_{I}} \, d\mathbf{r} \, d\mathbf{r}' .
\label{eq:force_total_term_RPA}
\end{aligned}
\end{equation}
This contribution needs to be added to the standard Hellmann-Feynman expression given in Eq.~(\ref{eq:force_hft_final}). 
The exact-exchange part has already been evaluated in Ref. \cite{contant2024OptimizedEffectivePotential} and shown to give a crucial contribution to the force. The RPA correlation contribution is readily obtained from Eq.~(\ref{eq:energy_correlation_rpa_develop}) as
\begin{equation}
\begin{aligned}
\displaystyle   \frac{\partial E_{\text{c}}^{\text{RPA}}}{\partial \mathbf{R}_{I}} = &  \int \limits_{0}^{\infty} \frac{du}{2\pi}\, \sum_{\alpha}^{N_{\alpha}} \frac{ - a_{\alpha}(iu) }{1 - a_{\alpha}(iu)} \\
& 
\times\iint \!\frac{\delta a_{\alpha}(iu)}{\delta v_{\text{eff}}(\mathbf{r},\mathbf{r}')} \frac{\partial v_{\text{NL}}(\mathbf{r}, \mathbf{r}')}{\partial \mathbf{R}_{I}} \, d\mathbf{r} \, d\mathbf{r}'.
\label{eq:force_fourth_term_RPA_developed}
\end{aligned}
\end{equation}

The collection of eigenvalues $\{a_{\alpha}\}$ and their variations are obtained directly from the calculation of the RPA correlation energy and potential. The detailed expression for $\delta a_{\alpha}(iu)/\delta v_{\text{eff}}$ when $v_{\text{eff}}$ is local is given in Refs.~\cite{nguyen2014InitioSelfconsistentTotalenergy, pitts2025SelfconsistentRandomPhase}. For a nonlocal $v_{\text{eff}}$, as in the present case, off-diagonal components also need to be calculated. Fortunately, this does not cause any additional complication. As a result, implementing the calculation of $\partial a_{\alpha}(iu)/\partial \mathbf{R}_{I}$ in the existing scf implementation is relatively straightforward. The computation of the extra scf-RPA correlation force term is also convenient as it can be done together with the computation of the RPA correlation potential. Furthermore, since only the pseudopotential is involved in Eq.~(\ref{eq:force_total_term_RPA}), the density and eigenvalue variations correspond to bare responses, which facilitates their evaluation. 
\subsection{Non-self-consistent forces}
By combining the standard HFT formulation of forces with the supplementary exchange and correlation terms identified and defined in Eq.~(\ref{eq:force_total_term_RPA}), we can calculate the RPA forces at self-consistency. Reaching self-consistency requires, however, around ten iterations in practice. In order to decrease the computational cost, it is motivated to explore the possibility of evaluating the forces from the non-self-consistent (nscf) RPA total energy.  

The force expression in Eq.~(\ref{eq:force_total_developed}) is valid for any KS starting point based on a local xc potential. For ease of discussion, we will hereafter focus on a particular starting point, the PBE approximation \cite{Ernzerhof1996,Ernzerhof1997}. In this case, the second term in Eq.~(\ref{eq:force_total_developed}) is evaluated with the PBE density, and the third force term is evaluated with the PBE xc potential. The derivative of $n^{\text{PBE}}$ with respect to $\mathbf{R}_{I}$ is determined from the first-order variation of the occupied PBE orbitals, which can be computed from the self-consistent Sternheimer equation,
\begin{equation}
\displaystyle \left( \hat{H}_{\rm KS} - \varepsilon_{\nu} \right) \Delta \phi_{\nu} = - \Delta \hat{V} \phi_{\nu},
\label{eq:sternheimer_equation_force}
\end{equation}
where $\hat{H}_{\rm KS}$ stands for the PBE KS Hamiltonian. For a given occupied band $\nu$, the quantity $\varepsilon_{\nu}$ corresponds to its KS eigenvalue, while $\Delta \phi_{\nu}$ is its first-order response to the applied perturbation, $\Delta \mathbf{R}_{I}$. The quantity $\Delta \hat{V}$ stands for the response of the KS potential to the same perturbation. We note that, since self-consistency is required, the only feasible starting points are those from LDA and GGA functionals, for which the DFPT formalism is implemented within \texttt{QE} \cite{baroni2001PhononsRelatedCrystal}.

For determining the fourth force term in Eq.~(\ref{eq:force_total_developed}), we again exploit the expression for $\delta E_{\text{xc}}^{\text{RPA}}/\delta  v_{\text{eff}}$ as in Eqs.~(\ref{eq:force_fourth_term_RPA})-(\ref{eq:force_fourth_term_RPA_developed}) but, this time, we include the variation of the full KS potential. Thus, we need to replace the third and fourth terms in Eq.~(\ref{eq:force_total_developed}) with the following expression
\begin{equation}
\begin{aligned}
\displaystyle \Delta \mathbf{F}_{\text{xc}}^{\text{nscf-RPA}} =\\
& \!\!\!\!\!\!\!\!\!\!\!\!\!\!\!\!\!\!\!\!\!\!\!\!\!\!\iiint \!v^{\text{PBE}}_{\text{xc}}(\mathbf{r}) \frac{\delta n^{\text{PBE}}(\mathbf{r})}{\delta v_{\text{eff}}(\mathbf{r}',\mathbf{r}'')} \frac{\partial v_{\text{eff}}(\mathbf{r}',\mathbf{r}'')}{\partial \mathbf{R}_{I}} \, d\mathbf{r} \, d\mathbf{r}' \, d\mathbf{r}''\\
& \!\!\!\!\!\!\!\!\!\!\!\!\!\!\!\!\!\!\!\!\!\!\!\!\!\! -\iint \! \frac{\delta E_{\text{xc}}^{\text{RPA}}}{\delta  v_{\text{eff}}(\mathbf{r},\mathbf{r}')} \frac{\partial v_{\text{eff}}(\mathbf{r},\mathbf{r}')}{\partial \mathbf{R}_{I}} \, d\mathbf{r} \, d\mathbf{r}' ,
\label{eq:nscf_force_fourth_term_RPA}
\end{aligned}
\end{equation}
where
\begin{equation}
\begin{aligned}
\displaystyle \frac{\partial v_{\text{eff}}(\mathbf{r},\mathbf{r}')}{\partial \mathbf{R}_{I}} =&\left\{\frac{\partial v_{\text{H}}(\mathbf{r})}{\partial \mathbf{R}_{I}}+\frac{\partial v^{\rm PBE}_{\text{xc}}(\mathbf{r})}{\partial \mathbf{R}_{I}}+\frac{\partial v_{\text{L}}(\mathbf{r})}{\partial \mathbf{R}_{I}}\right\}\\
&\times\delta(\mathbf{r}-\mathbf{r'})+\frac{\partial v_{\text{NL}}(\mathbf{r},\mathbf{r}')}{\partial \mathbf{R}_{I}}.
\label{eq:nscf_force_potential_term_RPA}
\end{aligned}
\end{equation}

The exact-exchange contribution is evaluated via the density matrix, which, similarly to the density, involves the self-consistent orbital responses to the perturbation $\Delta \mathbf{R}_{I}$ \cite{contant2024OptimizedEffectivePotential}. 

To summarize, the nscf-RPA forces are conveniently computed using the formalism of DFPT. The main difference with respect to the evaluation of scf-RPA forces is the requirement of a self-consistent evaluation of the Sternheimer equation. 
\begin{figure}[t]
\begin{center}
\includegraphics[scale=0.48,angle=0]{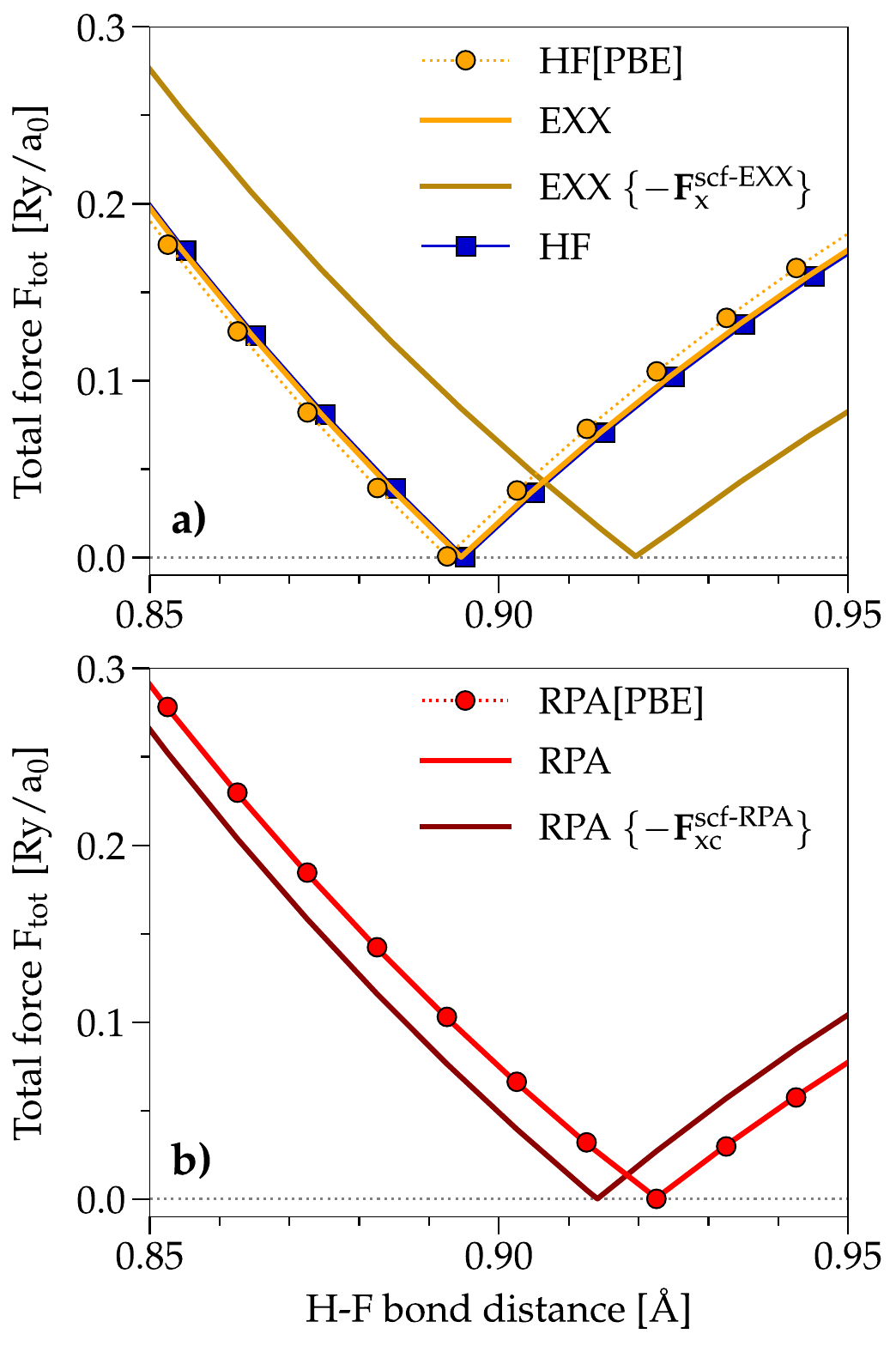}
\end{center}
\caption{\label{fig:Forces_FH_Scan} Change in the total force for the fluorane molecule according to \textbf{a)} HF, HF[PBE], EXX, and the uncorrected EXX ($-{\bf F}^{\rm scf\text{-}EXX}_{\rm x}$), \textbf{b)} RPA, RPA[PBE], and the uncorrected RPA ($-{\bf F}^{\rm scf\text{-}RPA}_{\rm xc}$).}
\end{figure}
\section{Implementation and testing}

We have implemented the different types of force calculations presented in the previous section within the \texttt{ACFDT} external package of the \texttt{QE} distribution \cite{giannozzi2009QUANTUMESPRESSOModular,giannozzi2017AdvancedCapabilitiesMaterials,giannozzi2020QuantumESPRESSOExascale,carnimeo2023QuantumESPRESSOOne}. Throughout this work, we have employed nonlocal pseudopotentials of the optimized norm-conserving Vanderbilt (ONCV) type \cite{hamann2013OptimizedNormconservingVanderbilt}. In this section, we will investigate the numerical accuracy achieved within scf-RPA and nscf-RPA. For the nscf calculations, we will always start from PBE. These calculations will be denoted RPA[PBE]. The scf calculations will simply be denoted RPA, without the brackets.

In order to judge whether the quality of the forces is sufficient for relaxing geometries using, for example, the Broyden-Fletcher-Goldfarb-Shanno (BFGS) algorithm \cite{Broyden1970,Fletcher1970,Goldfarb1970,Shanno1970}, we have considered the fluorane molecule, HF. The hydrogen atom has no core electron and local PPs are found to converge well. On the other hand, fluoride has two core electrons and, as such, benefit from using nonlocal PPs. 

A convenient quantity for monitoring the convergence of a geometry relaxation is the total force, defined as
\begin{equation}
\text{F}_{\text{tot}} = \sqrt{\sum_{I=1}^{M} \sum_{\beta=1}^{3} \left(F_{I}^{\beta}\right)^{2}},
\label{eq:Total_force_norm}
\end{equation}
where $M$ is the total number of atoms and $\beta$ spans the three spatial Cartesian coordinates. The total force takes into account the individual forces exerted on both the fluorine and hydrogen atoms, and should tend toward zero as the optimal geometry is approached. The change in the total force, calculated at different values of the H-F bond distance, is given in Fig.~\ref{fig:Forces_FH_Scan}, for different approximations. Panel a) focuses on the methods based on Hartree-Fock. 
The notation 'HF', also used to designate the fluorane molecule, refers to a self-consistent calculation done using the nonlocal Hartree-Fock potential, as implemented within the \texttt{PW} program of \texttt{QE}. The 'HF[PBE]' method corresponds to its non-self-consistent version that we have implemented within the \texttt{ACFDT} package, and 'EXX' refers to the self-consistent local version of HF that solves the OEP equation. Finally, 'EXX $\{-\mathbf{F}_\text{x}^\text{scf-EXX}\}$' corresponds to an EXX calculation that omits the extra exchange force contribution derived in Ref.~\cite{contant2024OptimizedEffectivePotential} (exchange part of Eq.~(\ref{eq:force_total_term_RPA})). For all these calculations, we used a kinetic energy cutoff of 100~Ry and a simulation cell of size 20~bohrs. For the HF and EXX calculations, a tight self-consistency threshold was set in order to ensure a good convergence of the results. The same was done for obtaining the self-consistent solution to the Sternheimer equation, necessary for computing the forces at the HF[PBE] level (see Eq.~(\ref{eq:sternheimer_equation_force})).

The HF molecule is found to have an equilibrium bond distance of 0.8924~\!\AA~using HF[PBE], 0.8948~\!\AA~for EXX, and 0.8952~\!\AA~for HF. The minimum in $\text{F}_{\text{tot}}$ is found, for all three methods, at the same position as the minimum in total energy, as expected. However, when we omit the extra exchange force term due to the nonlocal PP, the optimal bond length is about 0.03~\!\AA~larger than expected, with an overall error in the total force of about 0.085~Ry/a$_0$. When inspecting the individual forces exerted on each atom along a given Cartesian direction, the weight of the extra PP term is found, as expected, larger for the fluorine atom than the hydrogen, 0.11~Ry/a$_0$ against 1$\times$10$^{-4}$~Ry/a$_0$, respectively. 

Panel b) in Fig.~\ref{fig:Forces_FH_Scan} focuses on RPA using the same computational setup as for the EXX calculations. For the evaluation of the correlation terms, the frequency integration was performed on a grid of 8 imaginary frequency points, and 10 eigenvalues per valence electron ($N_{\rm eig}$=80) of the KS response function were determined (see Eq.~(\ref{eq:energy_rpa_eigenvalue_problem})). The bond distance is found at 0.9223~\!\AA~with RPA[PBE] and 0.9224~\!\AA~with RPA. In both cases, the minimum in the total force matches the minimum associated with the total energy, as expected. When removing the extra xc force term  (RPA $\{-\mathbf{F}_\text{xc}^\text{scf-RPA}\}$), the bias in forces leads to a shorter H-F bond. The error in the evaluation is about 0.01~\!\AA, which corresponds to a shift of 0.027~Ry/a$_0$ in the total force. Again, the extra xc force term is found to be the largest for F, about 0.04~Ry/a$_0$, while it is negligible for H. 

\begin{figure}[ht!]
\begin{center}
\includegraphics[scale=0.50,angle=0]{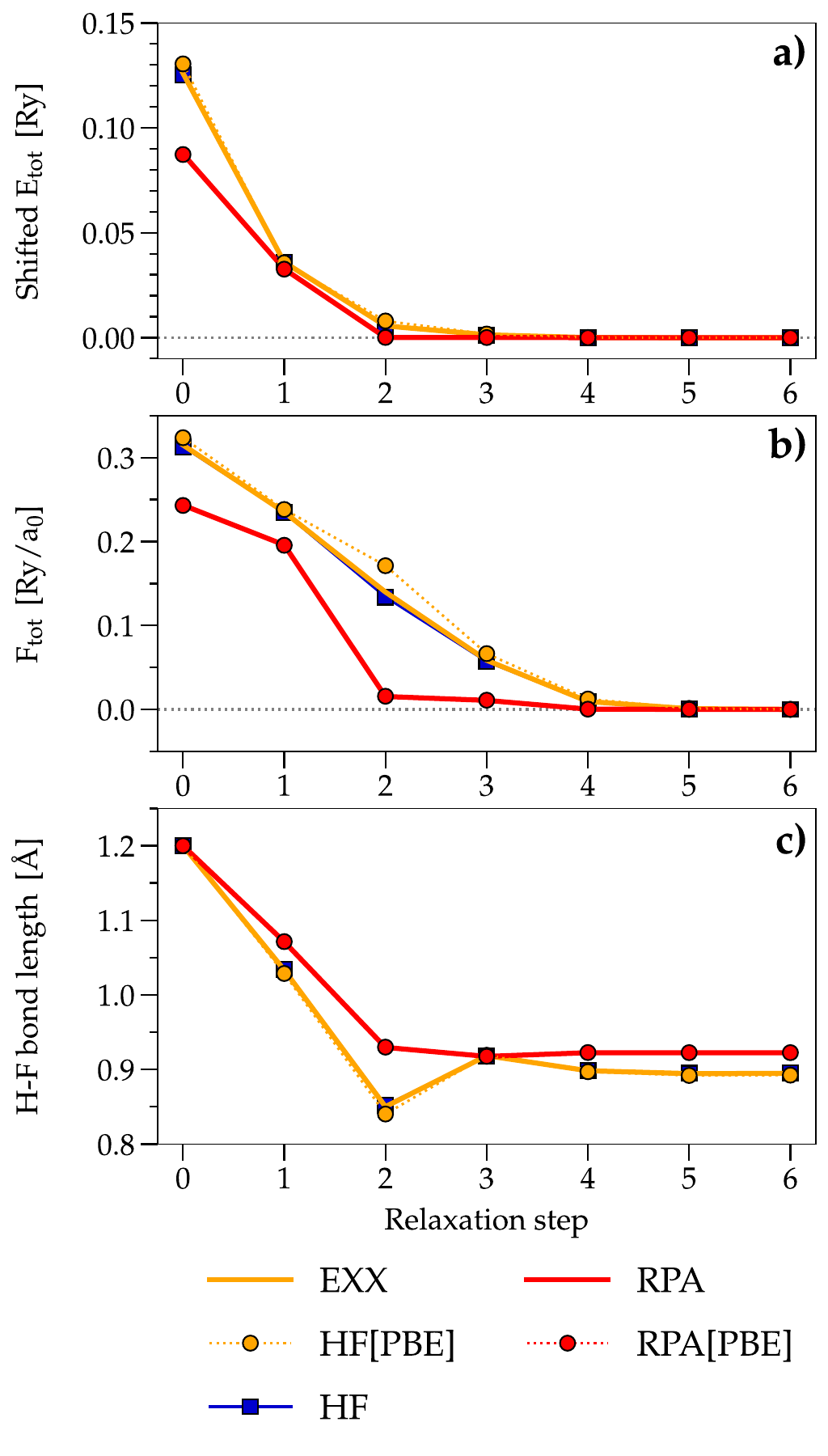}
\end{center}
\caption{\label{fig:Forces_FH_Relax} Evolution of \textbf{a)} the total energy, \textbf{b)} the total force, and \textbf{c)} the bond distance during the BFGS optimization procedure applied on the HF molecule.}
\end{figure}

Given the good agreement found at equilibrium geometry using either the forces or the total energy, we then investigated how well our calculated forces are exploited by the BFGS algorithm to relax the geometry of the fluorane molecule. 
For this test, the starting H-F bond distance was set to 1.2~\!\AA. During the relaxation procedure, we fixed the position of the hydrogen atom and only let the position of F to vary. The change over the relaxation steps in the total energy, the total force, and the H-F bond distance, are presented in panels \textbf{a)}, \textbf{b)}, and \textbf{c)} of Fig.~\ref{fig:Forces_FH_Relax}, respectively.

\begin{figure*}[ht!]
\begin{center}
\includegraphics[scale=0.55,angle=0]{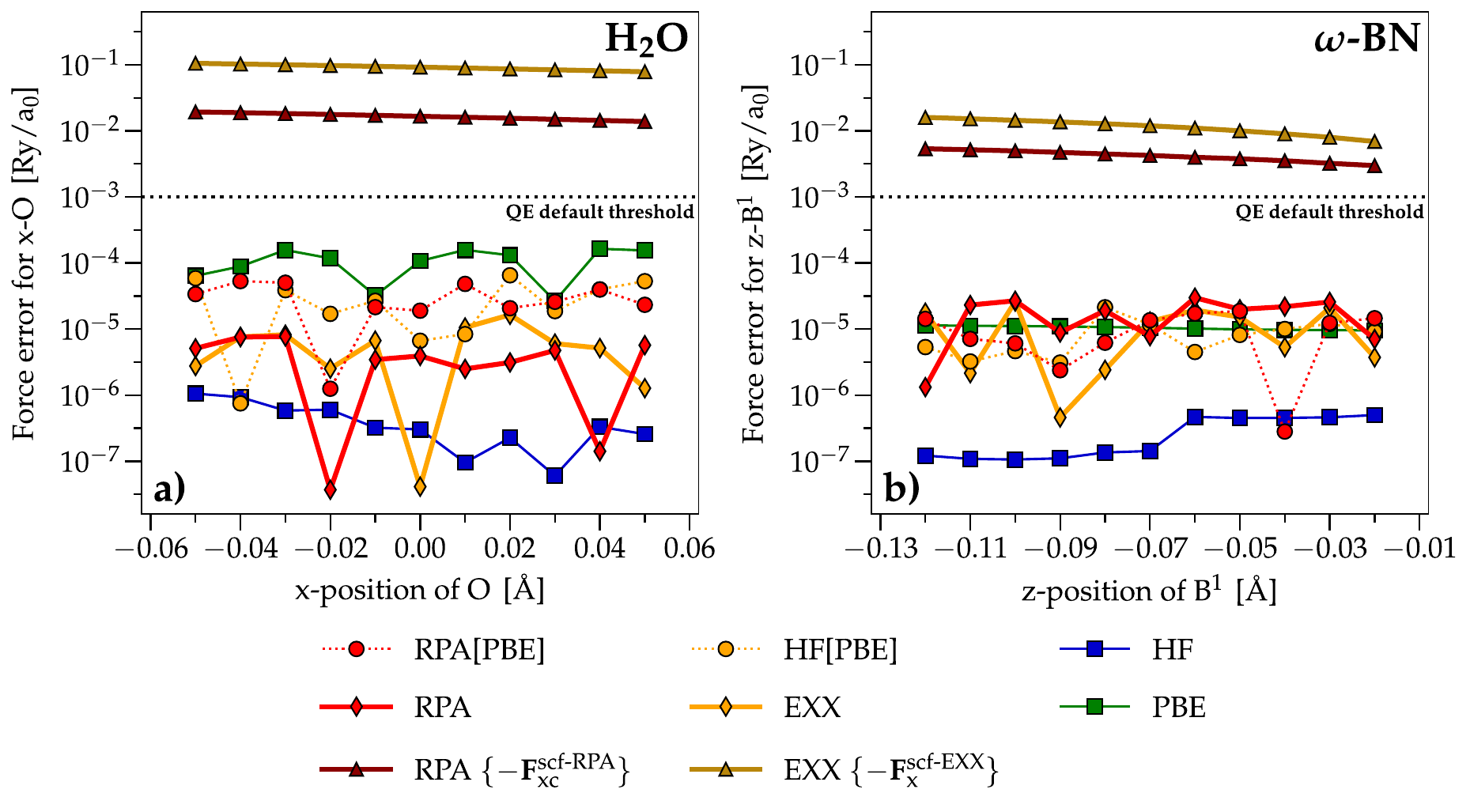}
\end{center}
\caption{\label{fig:Forces_H2O_BN_RPA} Estimated error for the value of the force exerted on \textbf{a)} a hydrogen atom (labeled H$^1$) along the O-H$^1$ bond of the water H$_2$O molecule and \textbf{b)} a boron ion (labeled B$^1$) along the z-direction ($\mathbf{c}$ crystal axis) of the $\omega$-BN crystal. The evolution is monitored with respect to changes in their position, using different approximations.}
\end{figure*}

In panel \textbf{a)}, the total energy is found to decrease during the relaxation, for every method considered. After only two to three BFGS steps, the variation in $\text{E}_{\text{tot}}$ becomes minimal. This evolution correlates well with the total force, which also reduces over the relaxation steps in panel \textbf{b)}. While the $\text{F}_{\text{tot}}$ indicator is initially large, its value decreases significantly for RPA methods after only two BFGS iterations. The decay is more continuous for EXX and HF[PBE] but agrees well with the HF calculations. In the \texttt{QE} code, the default convergence threshold for the geometry is set to 0.0001 Ry on the total energy, and 0.001~Ry/a$_0$ on each of the 3$M$ individual forces. In the situation considered here, the threshold on the force is reached at the 4th iteration for the RPA-based methods, and at the 5th BFGS step for the HF-based methods. This variation agrees well with the bond distance changes visible in panel c). The H-F bond length is found converged within 0.0005~\AA~from the 4th relaxation step for RPA[PBE] and RPA, and the 5th iteration for the exchange-only methods. This analysis confirms that our scf and nscf force implementation allows for a smooth optimization of the geometry using the BFGS algorithm.

Given the good quality found for our forces on this first test, we have then proceeded to determine their quantitative accuracy by comparing our analytical force calculations to the total energy finite-difference estimations. For this analysis, two different systems have been considered: the H$_2$O molecule and the wurtzite phase of the boron nitride solid ($\w$-BN, space group $P6_3/mmc$). For the H$_2$O molecule, calculations were performed using a 120~Ry cutoff for the kinetic energy, and 20~bohrs for the simulation cell size. With these settings, the PBE structure and PBE vibrational frequencies were well converged when using the forces. For the RPA calculations, we used the same plane-wave cutoff and cell size. For the evaluation of the correlation components, we used 10 eigenvalues per valence electron ($N_{\rm eig}$=80), and 12 imaginary frequency points.
Starting from the RPA equilibrium geometry (see Tab. I), we then varied one of the O-H bond length around 0.96~\AA, in steps of 0.01~\AA. The total energy finite-difference estimations were performed using the highly accurate five-point stencil derivation formula. 
For the $\w$-BN solid, we first optimized the unit cell using the PBE functional and then displaced one B$^{3+}$ cation along the $\mathbf{c}$ crystal axis. For these test calculations, we used a cutoff of 100~Ry, and a 2$\times$2$\times$2 \textbf{k}-point grid. For RPA, we used a grid made of 10 imaginary frequency points, determined 8 eigenvalues per valence electron ($N_{\rm eig}$=128), and used a 2$\times$2$\times$2 \textbf{q}-point grid. The error analyses for H$_2$O and $\omega$-BN are presented in panels a) and b) of Fig.~\ref{fig:Forces_H2O_BN_RPA}, respectively.

In panel a), we note that the largest error on the force is found for PBE, with a value of $2 \!\times\! 10^{-4}$~Ry/bohr, substantially above any of the other approximations considered. This error reduces, however, by increasing the energy cutoff, either on the plane waves or the density. At the level of HF, we instead note an almost perfect agreement between the analytical evaluation of the force and the numerical estimate from the total energy, with a difference between the two predictions below $10^{-6}$~Ry/bohr. The EXX method displays, at most, an error of $1.6 \!\times\! 10^{-5}$~Ry/bohr across the whole range of distances monitored. As discussed in Ref.~\cite{contant2024OptimizedEffectivePotential}, the oscillations and the overall error can be dampened further by increasing the kinetic energy cutoff and by tightening the self-consistent convergence conditions. 
If we compare the effect of self-consistency in the calculation, we see that the error found at the HF[PBE] level is larger than for HF and EXX. Despite this, the overall agreement remains of good quality, with a maximum error below $6.5 \!\times\! 10^{-5}$~Ry/bohr. A similar evolution is observed for the RPA method. At self-consistency, the error on RPA forces is found below $1 \!\times\! 10^{-5}$~Ry/bohr, similar to EXX. Without self-consistency, the evolution of RPA[PBE] matches HF[PBE], with a maximum error of $5.3 \!\times\! 10^{-5}$~Ry/bohr. Overall, the errors found are sufficiently small for exploiting the forces in a geometry relaxation procedure, given the 0.001~Ry/a$_0$ default threshold set on each individual force within \texttt{QE}. 

In panel b) of Fig.~\ref{fig:Forces_H2O_BN_RPA}, we find for $\w$-BN very similar results to the water molecule. The best accuracy is again found using HF, with a value on the order of 10$^{-7}$ Ry/a$_{0}$. The other methods are all gathered on the order of 10$^{-5}$ Ry/a$_{0}$. Unlike the case of the H$_2$O molecule, the scf and nscf evaluations of the RPA and EXX forces are found to return a similar accuracy. This analysis confirms the numerical accuracy of our force implementation, for both molecules and periodic solids.

Finally, in both panels a) and b), we note that if we omit the extra xc force terms (Eq. (\ref{eq:force_total_term_RPA})) for EXX and RPA, the methods denoted EXX $\{-{\bf F}^{\rm scf\text{-}EXX}_{\rm x}\}$ and RPA $\{-{\bf F}^{\rm scf\text{-}RPA}_{\rm x}\}$, respectively, the error in the analytical evaluation of the force is as large as 0.1-0.01 Ry/a$_{0}$. This further confirms the importance of including these terms for any approximation that relies on the OEP method.

\section{Results and discussion}
\label{secfive}
Now that we have verified and confirmed the high numerical precision of our scf and nscf RPA force calculations, we can exploit our implementation on a variety of systems, and assess the accuracy of the RPA for predicting geometries and vibrational frequencies within the harmonic approximation. 

Throughout this work, the vibrational frequencies are evaluated with the Phonopy software \cite{togo2023FirstprinciplesPhononCalculations,togo2023ImplementationStrategiesPhonopy}, which calculates the interatomic force constants by finite difference of forces before diagonalizing the dynamical matrix to obtain the normal modes of the system and their frequency. 

An implementation of the analytical forces for the approximations that include the exact-exchange kernel (RPAx, RPAx(1), and AC-SOSEX) is still missing. Nevertheless, for relatively simple systems, forces can still be evaluated using a total energy finite-difference scheme. For the systems studied here, this can be done on the normal mode coordinates, since they are insensitive to the choice of functional. Although more cumbersome, we have employed this approach within RPAx in order to assess the effect of the exact-exchange kernel on the calculated geometries and vibrational frequencies.

Below we present our results for a set of molecules (H$_2$, LiH, HF, N$_2$, H$_2$O, NH$_3$, and CH$_4$), and periodic solids (C, Si, and Ge).

\begin{figure}[t!]
\begin{center}
\includegraphics[scale=0.43,angle=0]{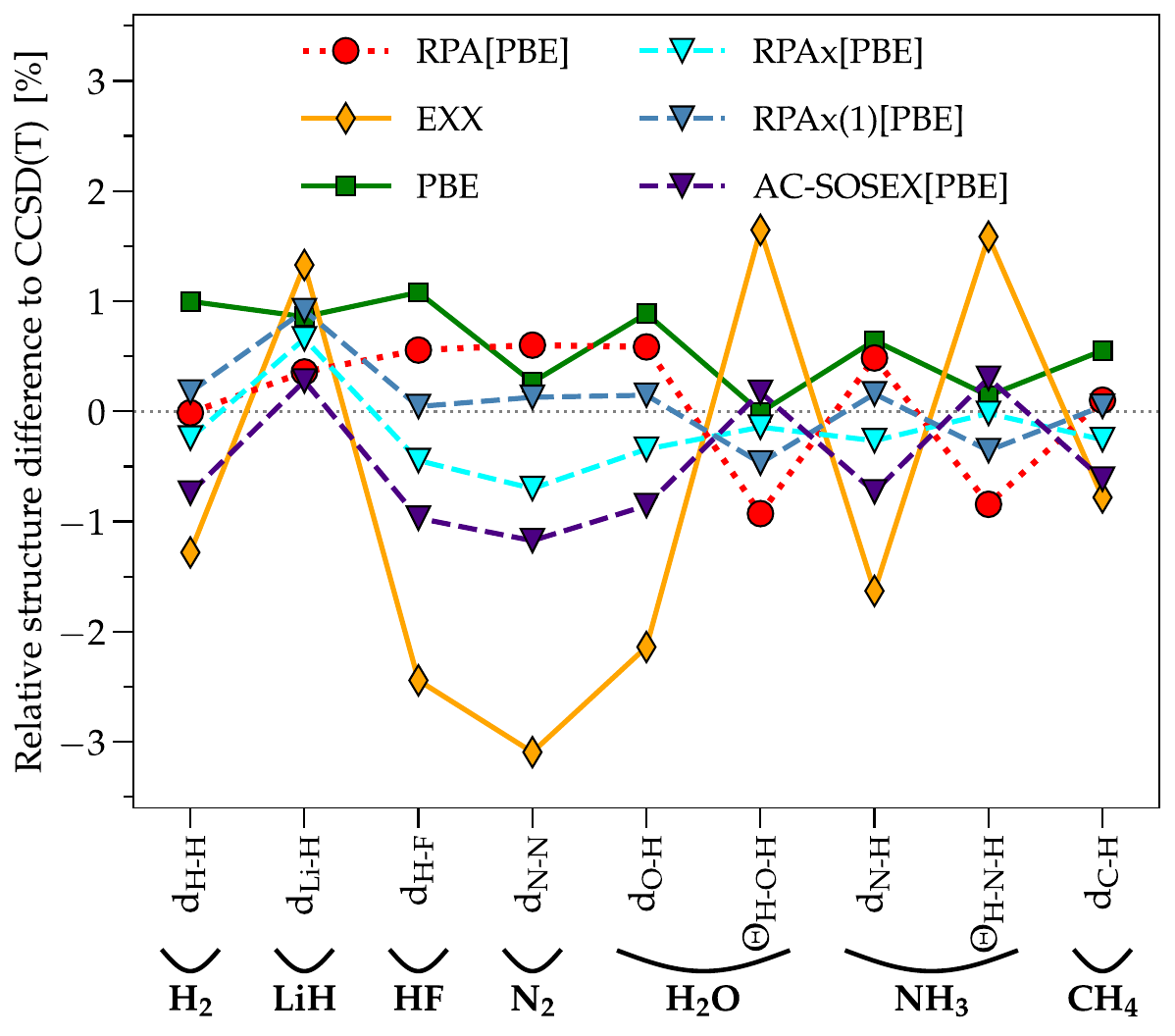}
\end{center}
\caption{\label{fig:Recap_Molecules_geometry} Relative error with respect to CCSD(T) for the structural parameters (bond distance, bond angle) listed in Table~\ref{tab:Tab_Molecules_Structure_RPA}.}
\end{figure}
\begin{figure}[t!]
\begin{center}
\includegraphics[scale=0.43,angle=0]{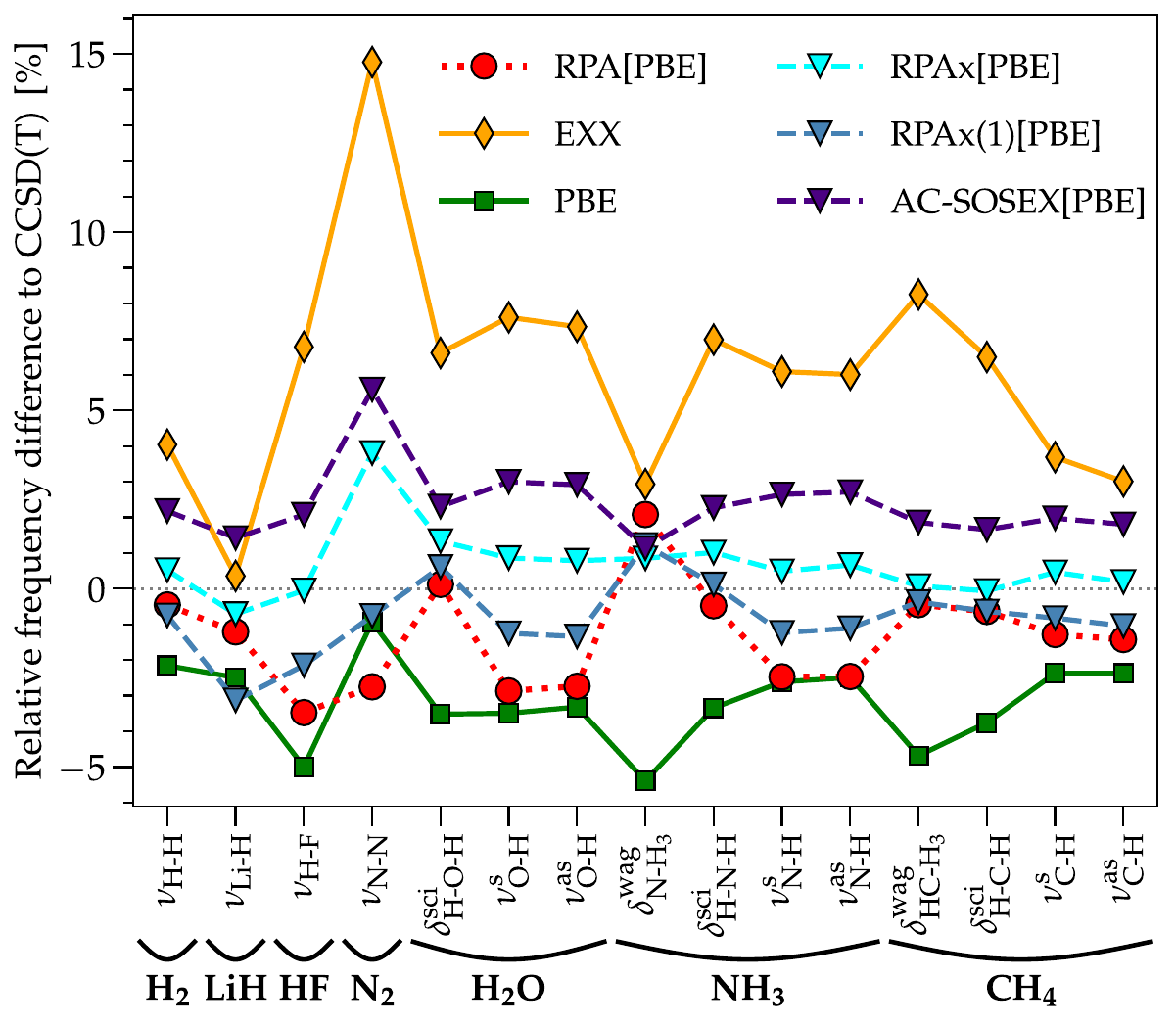}
\end{center}
\caption{\label{fig:Recap_Molecules_frequency} Relative error with respect to CCSD(T) for the various harmonic vibrational frequencies listed in Table~\ref{tab:Tab_Molecules_Frequency_RPA}.}
\end{figure}
\subsection{Molecules}
\label{secfiveA}
We have evaluated the performance of the ACFDT-based methods with respect to HF, PBE, and reference CCSD(T) calculations on a set of covalently bonded molecules: H$_2$, LiH, HF, N$_2$, H$_2$O, NH$_3$, and CH$_4$. For every system considered, a kinetic energy cutoff of 100~Ry has been used, and a cubic simulation cell of dimensions 20~bohrs has been used except for LiH, N$_2$, and CH$_4$, for which 22~bohrs was used instead. The correlation energy of the ACFDT-based methods was evaluated with a total of 8 imaginary frequency points, and 20 eigenvalues per valence electron for H$_{2}$ ($N_{\rm eig}$=40), 40 for LiH ($N_{\rm eig}$=80), and 10 for every remaining system ($N_{\rm eig}$=80 for HF, H$_2$O, NH$_3$, CH$_4$, and $N_{\rm eig}$=100 for N$_2$). Results obtained for the structural properties are presented in Tab.~\ref{tab:Tab_Molecules_Structure_RPA}, and the harmonic vibrational frequencies are presented in Tab.~\ref{tab:Tab_Molecules_Frequency_RPA}. The mean absolute percentage error (MAPE) indicated in these two tables is determined with respect to selected reference CCSD(T) results. They are illustrated in Figs.~\ref{fig:Recap_Molecules_geometry} and \ref{fig:Recap_Molecules_frequency}. %

\begin{table*}[ht!]
\caption{\label{tab:Tab_Molecules_Structure_RPA} Structural parameters of various molecules using different approximations. The bond distance is given in angstrom (\AA)~and the bond angle in degree (°). The CCSD(T) reference results are taken from Ref.~\cite{kupka2007PolarizationConsistentCorrelationConsistentBasis} (H$_2$ molecule), Ref.~\cite{noga1995PerformanceExplicitlyCorrelated} (LiH molecule), and Ref.~\cite{tew2007BasisSetLimit} for the remaining systems.
}
\vspace{5pt}
\centering
\begin{tabular}{M{1.3cm} M{1.5cm} M{1.2cm} M{1.2cm} M{1.2cm} M{1.5cm} M{2.1cm} M{1.2cm} M{1.0cm} M{1.2cm} M{1.2cm} M{1.6cm}}
    \toprule
     & \textbf{Geometry} & \makecell{\textbf{RPA} \\ \textbf{(scf)}} & \makecell{\textbf{RPA} \\ \textbf{[PBE]}} & \makecell{\textbf{RPAx} \\ \textbf{[PBE]}} & \makecell{\textbf{RPAx(1)} \\ \textbf{[PBE]}} & \makecell{\textbf{AC-SOSEX} \\ \textbf{[PBE]}} & \makecell{\textbf{EXX} \\ \textbf{(scf)}} & \makecell{\textbf{HF} \\ \textbf{(scf)}} & \makecell{\textbf{HF} \\ \textbf{[PBE]}} & \makecell{\textbf{PBE} \\ \textbf{(scf)}} & \makecell{\textbf{CCSD(T)} \\ \text{Ref.}} \\
    \midrule
    \textbf{H$_{\mathbf{2}}$} & \textbf{d}$_{\textbf{H-H}}$ & 0.7414 & 0.7415 & 0.7398 & 0.7429 & 0.7361 & 0.7321 & 0.7321 & 0.7311 & 0.7490 & 0.7416 \\
    \textbf{LiH} & \textbf{d}$_{\textbf{Li-H}}$ & 1.6019 & 1.6017 & 1.6065 & 1.6106 & 1.6004 & 1.6172 & 1.6165 & 1.6099 & 1.6097 & \!\!\!1.596 \\
    \textbf{HF} & \textbf{d}$_{\textbf{H-F}}$ & 0.9224 & 0.9223 & 0.9131 & 0.9176 & 0.9083 & 0.8948 & 0.8952 & 0.8924 & 0.9271 & 0.9172 \\
    \textbf{N$_{\mathbf{2}}$} & \textbf{d}$_{\textbf{N-N}}$ & 1.1052 & 1.1054 & 1.0911 & 1.1002 & 1.0859 & 1.0648 & 1.0653 & 1.0626 & 1.1017 & 1.0988 \\
    \textbf{H$_{\mathbf{2}}$O} & \textbf{d}$_{\textbf{O-H}}$ & 0.9637 & 0.9638 & 0.9549 & 0.9596 & 0.9500 & 0.9377 & 0.9382 & 0.9359 & 0.9667 & 0.9582 \\
     & $\boldsymbol{\Theta}_{\textbf{H-O-H}}$ & 103.52 & 103.49 & 104.31 & 103.97 & 104.64 & 106.18 & 106.05 & 106.14 & 104.45 & 104.46 \\
    \textbf{NH$_{\mathbf{3}}$} & \textbf{d}$_{\textbf{N-H}}$ & 1.0168 & 1.0169 & 1.0093 & 1.0136 & 1.0046 & 0.9955 & 0.9959 & 0.9941 & 1.0185 & 1.0120 \\
    & $\boldsymbol{\Theta}_{\textbf{H-N-H}}$ & 105.74 & 105.71 & 106.59 & 106.23 & 106.93 & 108.30 & 108.13 & 108.28 & 106.76 & 106.61 \\
    \textbf{CH$_{\mathbf{4}}$} & \textbf{d}$_{\textbf{C-H}}$ & 1.0884 & 1.0886 & 1.0847 & 1.0880 & 1.0808 & 1.0790 & 1.0791 & 1.0777 & 1.0935 & 1.0875 \\
    
    \midrule
    \textbf{MAPE} & \textbf{all} & 0.49 & 0.50 & 0.34 & 0.27 & 0.65 & 1.77 & 1.71 & 1.83 & 0.60 & 0.00 \\
    \textbf{MAPE} & \textbf{bond} & 0.38 & 0.39 & 0.42 & 0.23 & 0.77 & 1.81 & 1.78 & 1.90 & 0.75 & 0.00 \\
    \textbf{MAPE} & \textbf{angle} & 0.86 & 0.89 & 0.08 & 0.41 & 0.24 & 1.62 & 1.47 & 1.59 & 0.08 & 0.00 \\
    \bottomrule
\end{tabular}
\end{table*}

\begin{table*}[ht!]
\caption{\label{tab:Tab_Molecules_Frequency_RPA} Harmonic vibrational frequencies (cm$^{-1}$) of various molecules using different approximations. The CCSD(T) reference results are taken from Ref.~\cite{kupka2007PolarizationConsistentCorrelationConsistentBasis} (H$_2$ molecule), Ref.~\cite{noga1995PerformanceExplicitlyCorrelated} (LiH molecule), and Ref.~\cite{tew2007BasisSetLimit} for the remaining systems.
}
\vspace{5pt}
\centering
\begin{tabular}{M{1.3cm} M{1.5cm} M{1.2cm} M{1.2cm} M{1.2cm} M{1.5cm} M{2.1cm} M{1.2cm} M{1.0cm} M{1.2cm} M{1.2cm} M{1.6cm}}
    \toprule
     & \textbf{Mode} & \makecell{\textbf{RPA} \\ \textbf{(scf)}} & \makecell{\textbf{RPA} \\ \textbf{[PBE]}} & \makecell{\textbf{RPAx} \\ \textbf{[PBE]}} & \makecell{\textbf{RPAx(1)} \\ \textbf{[PBE]}} & \makecell{\textbf{AC-SOSEX} \\ \textbf{[PBE]}} & \makecell{\textbf{EXX} \\ \textbf{(scf)}} & \makecell{\textbf{HF} \\ \textbf{(scf)}} & \makecell{\textbf{HF} \\ \textbf{[PBE]}} & \makecell{\textbf{PBE} \\ \textbf{(scf)}} & \makecell{\textbf{CCSD(T)} \\ \text{Ref.}} \\
    \midrule
    \textbf{H$_{\mathbf{2}}$} & $\boldsymbol{\nu}_{\text{\textbf{H-H}}}$ & 4385 & 4385 & 4428 & 4372 & 4501 & 4583 & 4583 & 4609 & 4310 & 4405 \\
    \textbf{LiH} & $\boldsymbol{\nu}_{\text{\textbf{Li-H}}}$ & 1388 & 1389 & 1396 & 1362 & 1426 & 1411 & 1412 & 1439 & 1371 & 1406 \\
    \textbf{HF} & $\boldsymbol{\nu}_{\text{\textbf{H-F}}}$ & 3996 & 3999 & 4141 & 4054 & 4229 & 4424 & 4414 & 4469 & 3936 & 4143 \\
    \textbf{N$_{\mathbf{2}}$} & $\boldsymbol{\nu}_{\text{\textbf{N-N}}}$ & 2298 & 2298 & 2453 & 2345 & 2495 & 2712 & 2706 & 2736 & 2341 & 2363 \\
    \textbf{H$_{\mathbf{2}}$O} & $\boldsymbol{\delta}_{\text{\textbf{H-O-H}}}^{\text{\textbf{sci}}}$ & 1650 & 1651 & 1671 & 1659 & 1687 & 1758 & 1761 & 1767 & 1591 & 1649 \\
     & $\boldsymbol{\nu}_{\text{\textbf{O-H}}}^{\text{\textbf{s}}}$ & 3728 & 3726 & 3869 & 3788 & 3951 & 4128 & 4117 & 4161 & 3702 & 3836 \\
     & $\boldsymbol{\nu}_{\text{\textbf{O-H}}}^{\text{\textbf{as}}}$ & 3841 & 3839 & 3978 & 3894 & 4062 & 4237 & 4224 & 4268 & 3816 & 3947 \\
    \textbf{NH$_{\mathbf{3}}$} & $\boldsymbol{\delta}_{\text{\textbf{N-H$_\mathbf{3}$}}}^{\text{\textbf{wag}}}$ & 1079 & 1079 & 1066 & 1070 & 1069 & 1088 & 1097 & 1090 & 1000 & 1057 \\
     & $\boldsymbol{\delta}_{\text{\textbf{H-N-H}}}^{\text{\textbf{sci}}}$ & 1667 & 1667 & 1692 & 1677 & 1713 & 1792 & 1794 & 1797 & 1619 & 1675 \\
     & $\boldsymbol{\nu}_{\text{\textbf{N-H}}}^{\text{\textbf{s}}}$ & 3397 & 3395 & 3498 & 3438 & 3573 & 3693 & 3685 & 3717 & 3390 & 3481 \\
     & $\boldsymbol{\nu}_{\text{\textbf{N-H}}}^{\text{\textbf{as}}}$ & 3526 & 3524 & 3637 & 3573 & 3711 & 3830 & 3818 & 3854 & 3523 & 3613 \\
    \textbf{CH$_{\mathbf{4}}$} & $\boldsymbol{\delta}_{\text{\textbf{HC-H$_\mathbf{3}$}}}^{\text{\textbf{wag}}}$ & 1340 & 1339 & 1346 & 1340 & 1370 & 1456 & 1459 & 1458 & 1282 & 1345 \\
    & $\boldsymbol{\delta}_{\text{\textbf{H-C-H}}}^{\text{\textbf{sci}}}$ & 1558 & 1560 & 1569 & 1560 & 1596 & 1672 & 1673 & 1676 & 1511 & 1570 \\
    & $\boldsymbol{\nu}_{\text{\textbf{C-H}}}^{\text{\textbf{s}}}$ & 2998 & 2997 & 3050 & 3011 & 3096 & 3148 & 3147 & 3164 & 2964 & 3036 \\
    & $\boldsymbol{\nu}_{\text{\textbf{C-H}}}^{\text{\textbf{as}}}$ & 3114 & 3113 & 3164 & 3125 & 3215 & 3253 & 3247 & 3274 & 3083 & 3158 \\
    \midrule
    \textbf{MAPE} & \textbf{all} & 1.65 & 1.66 & 0.79 & 1.10 & 2.37 & 6.07 & 6.04 & 6.75 & 3.20 & 0.00 \\
    \textbf{MAPE} & \textbf{stretch} & 2.10 & 2.11 & 0.85 & 1.36 & 2.63 & 5.97 & 5.79 & 6.86 & 2.72 & 0.00 \\
    \textbf{MAPE} & \textbf{bend} & 0.75 & 0.75 & 0.67 & 0.59 & 1.84 & 6.26 & 6.54 & 6.54 & 4.14 & 0.00 \\
    \bottomrule
\end{tabular}
\end{table*}
\begin{table*}[ht!]
\caption{\label{tab:Tab_Diamond_alat} The lattice parameter $a$ (\AA) of diamond. The DMC results are taken from Ref.~\cite{maezono2007EquationStateRaman}, and the LRDMC results from Ref.~\cite{nakano2021AtomicForcesQuantum}. For the experimental value, the effect of the zero-point vibrational energy and temperature has been subtracted (indicated with a $\dagger$ symbol) \cite{harl2010AssessingQualityRandom}. The measured experimental lattice parameter is 3.567~\AA~\cite{occelli2003PropertiesDiamondHydrostatic}.
}
\vspace{5pt}
\centering
\begin{tabular}{M{1.5cm} M{1.2cm} M{2.1cm} M{2.1cm} M{1.0cm} M{1.6cm} M{1.8cm} M{1.4cm}}
    \toprule
     \makecell{\textbf{RPA} \\ \textbf{[PBE]}} & \makecell{\textbf{RPAx} \\ \textbf{[PBE]}} & \makecell{\textbf{PBE0(17\%)} \\ \textbf{(scf)}} & \makecell{\textbf{PBE0(34\%)} \\ \textbf{(scf)}} & \makecell{\textbf{PBE} \\ \textbf{(scf)}} & \makecell{\textbf{DMC} \\ \text{Ref.~\cite{maezono2007EquationStateRaman}}} & \makecell{\textbf{LRDMC} \\ \text{Ref.~\cite{nakano2021AtomicForcesQuantum}}} & \makecell{\textbf{Expt.} \\ \text{Ref.~\cite{harl2010AssessingQualityRandom}}}
     \\
    \midrule
    3.567 & 3.552 & 3.552 & 3.536 & 3.569 & \,\,\,\,\,\,3.563$\pm$2 & \,\,\,\,\,\,\,3.547$\pm$1 & \,\,3.553$^{\dagger}$ \\
    \bottomrule
\end{tabular}
\end{table*}

\begin{table*}[ht!]
\caption{\label{tab:Tab_Diamond_Frequency} The harmonic optical $\Gamma$-phonon frequency $\nu$ (cm$^{-1}$) of diamond. The DMC result is taken from Ref.~\cite{maezono2007EquationStateRaman}, and the LRDMC result from Ref.~\cite{nakano2021AtomicForcesQuantum}. For every method, the frequency has been calculated using the experimental lattice parameter   (3.567~\AA~\cite{occelli2003PropertiesDiamondHydrostatic}). The anharmonic correction has been estimated to -17.3~cm$^{-1}$ in Ref.~\cite{vanderbilt1986CalculationAnharmonicPhonon}.
}
\vspace{5pt}
\centering
\begin{tabular}{M{1.5cm} M{1.2cm} M{1.5cm} M{2.0cm} M{2.1cm} M{2.1cm} M{1.0cm} M{1.6cm} M{1.8cm} M{1.4cm}}
    \toprule
     \makecell{\textbf{RPA} \\ \textbf{[PBE]}} & \makecell{\textbf{RPAx} \\ \textbf{[PBE]}} & \makecell{\textbf{RPAx(1)} \\ \textbf{[PBE]}} & \makecell{\textbf{AC-SOSEX} \\ \textbf{[PBE]}} & \makecell{\textbf{PBE0(17\%)} \\ \textbf{(scf)}} & \makecell{\textbf{PBE0(34\%)} \\ \textbf{(scf)}} & \makecell{\textbf{PBE} \\ \textbf{(scf)}} & \makecell{\textbf{DMC} \\ \text{Ref.~\cite{maezono2007EquationStateRaman}}} & \makecell{\textbf{LRDMC} \\ \text{Ref.~\cite{nakano2021AtomicForcesQuantum}}} & \makecell{\textbf{Expt.} \\ \text{Ref.~\cite{occelli2003PropertiesDiamondHydrostatic}}}
     \\
    \midrule
    1342 & 1362 & 1348 & 1368 & 1331 & 1368 & 1290 & \,\,\,\,\,\,1375$\pm$4 & \,\,\,\,\,\,\,1384$\pm$8 & \,\,1334 \\
    \bottomrule
\end{tabular}
\end{table*}
First of all, we note that the results obtained with self-consistent RPA are very similar to the results from non-self-consistent RPA. The RPA[PBE] approach, therefore, represents an interesting alternative to fully self-consistent RPA given its cheaper computational cost. Although we do not have access to self-consistent RPAx results, we expect a very similar behavior for this method. For HF, we see slightly larger differences between EXX and EXX[PBE], most likely due to a larger dissimilarity between the EXX and PBE orbitals. Despite that, the differences in terms of both structural properties and vibrational frequencies remain very small.

If we now compare the performance of the different approximations with respect to the most recent CCSD(T) results, we see that RPA generally overestimates the bond distances and underestimates the vibrational frequencies of the "bond stretching" type. Nonetheless, the RPA method systematically improves the results when comparing it to PBE, lowering the MAPE from 2.72\% to 2.11\%, on the average. A larger improvement is seen for the vibrational modes of "bending" type, present in H$_2$O, NH$_3$, and CH$_4$. The RPA reduces the 4.14\% error in PBE to 0.75\%, despite the fact that the bond angles are slightly worse in RPA as compared to PBE. 
The molecular geometries predicted by our implementation are in good agreement with the all-electron RPA calculations previously reported in the literature using Gaussian basis sets \cite{thierbach2020AnalyticEnergyGradientsa,bates2025FrozenCoreAnalyticalGradients,tahir2025AnalyticalGradientsRandomPhase}.

Overall, the RPAx method displays an outstanding performance. Compared to RPA, it greatly improves the bond distances and bond stretching frequencies, yielding MAPE values as low as 0.42\% and 0.85\%, respectively. The quality of the bending modes remains of the same high quality as in RPA but with an improved description of the bond angles. 

In Tabs.~\ref{tab:Tab_Molecules_Structure_RPA} and \ref{tab:Tab_Molecules_Frequency_RPA}, we have included the results obtained with the two alternative partial resummations of RPAx, which were discussed in Sec. III. Overall, AC-SOSEX worsens the performance with respect to RPA, compressing the geometries and overestimating both types of vibrational frequencies. These results are consistent with a previous work that evidenced a failure of SOSEX to improve upon RPA potentials \cite{Vacondio2022}. 
The RPAx(1), on the other hand, produces results of close quality to RPAx. On the average, RPAx(1) structures appear further improved with respect to RPAx but the vibrational frequencies are slightly underestimated, with a MAPE of 1.1\%, which can be compared to the slightly overestimated frequencies within RPAx (MAPE: 0.79\%). The systematic improvement of RPAx and RPAx(1) over RPA clearly indicates that the inclusion of the exact-exchange kernel is of importance to refine the structural and vibrational properties of molecules.
\begin{figure}[b]
\begin{center}
\includegraphics[scale=0.37,angle=0]{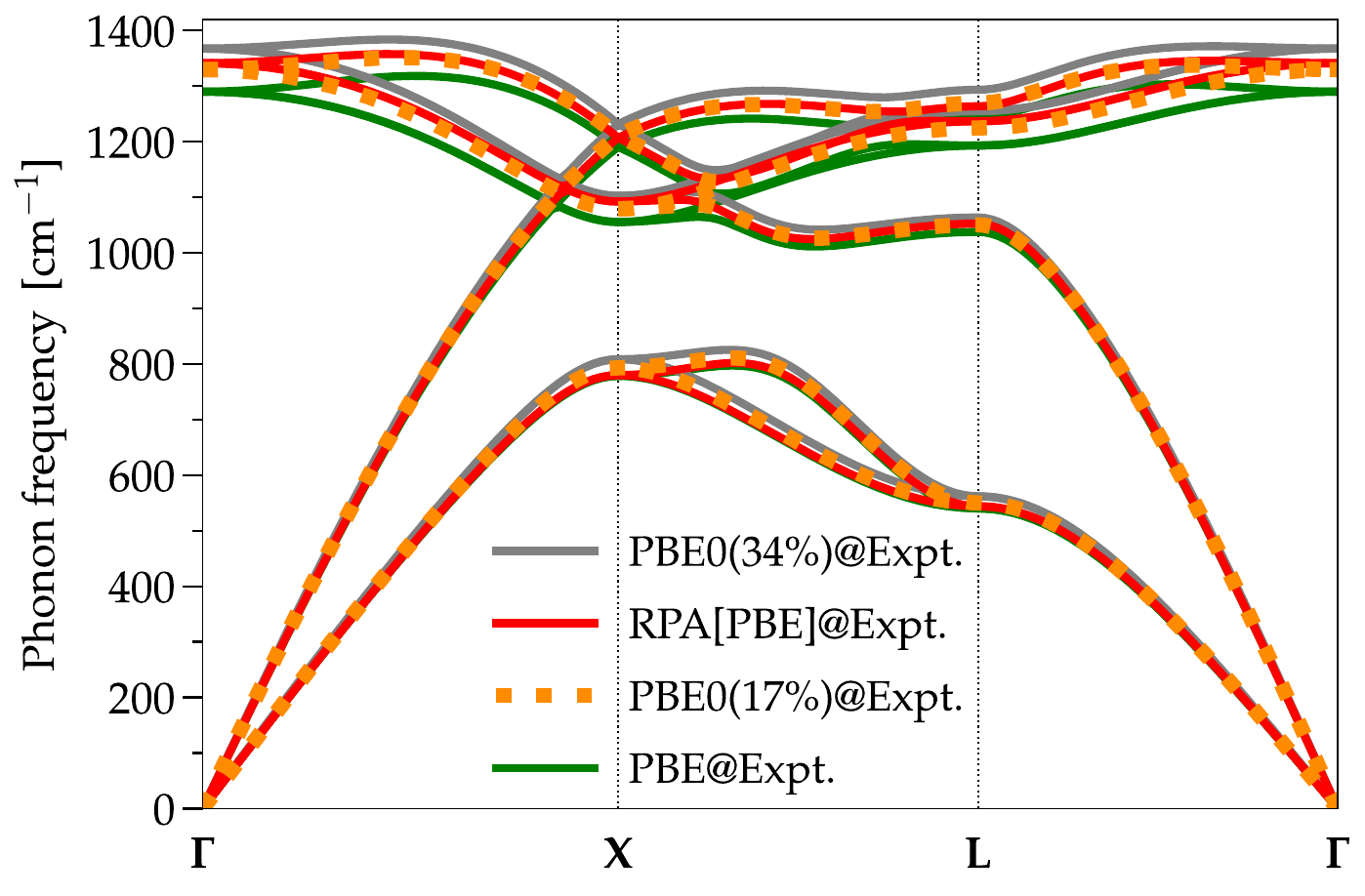}
\end{center}
\caption{\label{fig:Diamond_RPA_geom} Phonon dispersion of diamond, as calculated using different methods (PBE, PBE0(17\%), PBE0(34\%), RPA[PBE]) on the experimental unit cell (@Expt.).}
\end{figure}
\begin{table*}[t]
\caption{\label{tab:Tab_Silicon_Frequency} The $\Gamma$-point optical phonon frequency $\nu$ (cm$^{-1}$) of solid silicon, evaluated within the harmonic approximation. For every method, the frequency has been calculated using the experimental lattice parameter  (5.43~\AA~\cite{kittel2005IntroductionSolidState}). The anharmonic correction has been estimated to -3.5~cm$^{-1}$ in Ref.~\cite{vanderbilt1986CalculationAnharmonicPhonon}.
}
\vspace{5pt}
\centering
\begin{tabular}{M{2.0cm} M{2.0cm} M{2.0cm} M{2.0cm} M{2.1cm} M{2.1cm} M{2.0cm} M{1.6cm} M{2.8cm}}
    \toprule
     \makecell{\textbf{RPA} \\ \textbf{[PBE]}} & \makecell{\textbf{RPAx} \\ \textbf{[PBE]}} & \makecell{\textbf{RPAx(1)} \\ \textbf{[PBE]}} & \makecell{\textbf{AC-SOSEX} \\ \textbf{[PBE]}} & \makecell{\textbf{PBE0(29\%)} \\ \textbf{(scf)}} & \makecell{\textbf{PBE} \\ \textbf{(scf)}} &   
     \makecell{\textbf{Expt.}\\ Ref.~\cite{Kulda1994}}
     \\
    \midrule
    522 & 531 & 527 & 536 & 536 & 514 &523\\
    \bottomrule
\end{tabular}
\end{table*}

\subsection{Diamond, silicon, and germanium}
We will now study the performance of the different ACFDT methods on solids. Let us start with the carbon diamond phase. Despite the simplicity of this system, which only has one degree of freedom - its lattice parameter - an accurate theoretical determination of the optical phonon at $\Gamma$ is still missing. Calculations performed at the DMC level have predicted a harmonic frequency at 1374-1384 cm$^{-1}$ \cite{maezono2007EquationStateRaman,nakano2021AtomicForcesQuantum}, which overestimates the experimental frequency at 1334 cm$^{-1}$ \cite{occelli2003PropertiesDiamondHydrostatic}. If the accuracy of the DMC calculations is within the stochastic error bar (4-8 cm$^{-1}$), these results would suggest a quite large anharmonic renormalization, which disagrees with previous estimations done by total energy finite-difference calculations up to fourth order \cite{vanderbilt1984CalculationPhononPhononInteractions} and by PBE path integral molecular dynamics calculations \cite{nakano2021AtomicForcesQuantum}, which estimate a correction of -17.3 cm$^{-1}$ and -13.7 cm$^{-1}$, respectively.

Given this inconsistency, we have performed highly converged calculations with RPA, RPAx, RPAx(1), and AC-SOSEX at the harmonic level. 
In order to estimate the anharmonic shift as consistently as possible, yet in a computationally feasible way, we have optimized the fraction of exact exchange in the PBE0 hybrid functional, according to the procedure in Refs. \cite{hellgren2021RandomPhaseApproximation,pitts2025SelfconsistentRandomPhase}. In RPA we find a fraction of 17\% and in RPAx 34\%. 
The lattice parameter, calculated for some of the methods, is presented in Tab.~\ref{tab:Tab_Diamond_alat}. The harmonic optical $\Gamma$-phonon, calculated at the experimental lattice constant ($a$ = 3.567~\AA ~\cite{occelli2003PropertiesDiamondHydrostatic}), is presented in Tab.~\ref{tab:Tab_Diamond_Frequency}. 

We first note that the RPA lattice parameter is in seemingly perfect agreement with the experimental value. 
However, subtracting the zero-point vibrational energy (ZPE) contribution from the experimental extrapolated zero-temperature result gives $a$ = 3.553~\AA, as shown in Ref. ~\cite{harl2010AssessingQualityRandom}. By comparing this value to our theoretical estimates, we find instead RPAx[PBE] to be in excellent agreement with experiments. Looking at the other methods, we see that PBE is close to RPA, while the optimized hybrid functionals underestimate the lattice constant with respect to the prediction of the underlying RPA/RPAx method. 

Let us now investigate the optical phonon, calculated at the $\Gamma$-point with each method but at the measured experimental lattice constant. The phonon has been converged within 1-2~cm$^{-1}$ for every method. The ACFDT-based methods required a 80 Ry plane-wave cutoff, and a total of 80 eigenvalues and 8 imaginary frequency points for the correlation energy. For RPA/RPAx, we employed a 6$\times$6$\times$6/5$\times$5$\times$5 $\Gamma$-centered $\mathbf{k}$-point grid, and then extrapolated the last 2.5/8.5~cm$^{-1}$ using the PBE0 convergence curves. For the other methods, we used a plane-wave cutoff of 200 Ry and a 8$\times$8$\times$8 $\mathbf{k}$-point-grid. Convergence plots can be found in the Appendix. 
\begin{figure}[b]
\begin{center}
\includegraphics[scale=0.37,angle=0]
{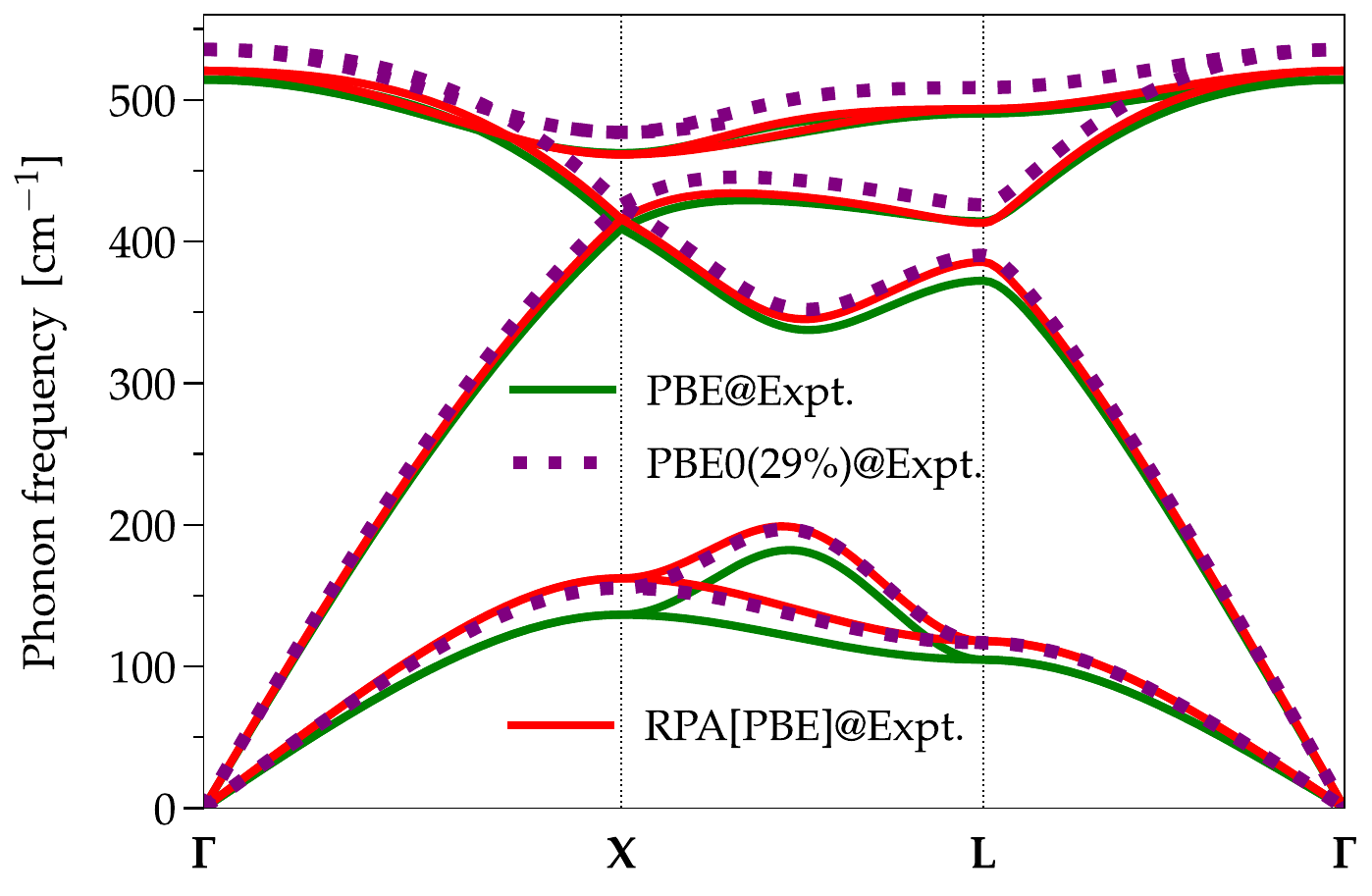}
\end{center}
\caption{\label{fig:Silicon_RPA_dispersion} Phonon dispersion of silicon, as calculated using different methods (PBE, PBE0(29\%), RPA[PBE]) on the experimental unit cell (@Expt.).}
\end{figure}

The impact of self-consistency turns out to be negligible in RPA. Self-consistent RPA gives a frequency 1~cm$^{-1}$ lower than RPA[PBE] in the 3$\times$3$\times$3 and 4$\times$4$\times$4 $\mathbf{k}$-point-grid calculations. We, therefore, did not pursue self-consistent RPA calculations with larger grids. Similarly, different starting points for RPAx (such as PBE and EXX) also showed very small differences in the frequency. 

The RPA predicts a harmonic frequency at 1342~cm$^{-1}$, which is in good agreement with Ref. \cite{ramberger2017AnalyticInteratomicForces}, and close to the experimental value at 1334~cm$^{-1}$. Adding the PBE anharmonic correction of -17~cm$^{-1}$, found in earlier works, gives us an experimental harmonic reference at 1351~cm$^{-1}$. The RPA thus underestimates experiments by 9~cm$^{-1}$, which is less than 1\%. This can be compared to the PBE error of 4.5\%, or 61~cm$^{-1}$. The harmonic RPAx[PBE] value is at 1362~cm$^{-1}$, which corresponds to an increase of 20~cm$^{-1}$ with respect to RPA, and to an 11~cm$^{-1}$ overestimation with respect to the harmonic experimental frequency.  Clearly, the RPA and RPAx calculations produce a phonon frequency in closer agreement with experiments than DMC, which overestimates by 20-40 ~cm$^{-1}$, taking into account error bars.
If we consider the optimized hybrid functionals, we see that PBE0(17\%) is only slightly below RPA, and PBE0(34\%) only slightly above RPAx. This agreement suggests that these methods work quite well in approximating the performance of the method that was used to optimize them.
\begin{table*}[t]
\caption{\label{tab:Tab_Germanium_Frequency} The $\Gamma$-point optical phonon frequency $\nu$ (cm$^{-1}$) of solid germanium, evaluated within the harmonic approximation. For every method, the frequency has been calculated using the experimental lattice parameter  (5.6575~\AA~\cite{straumanis1952LatticeParametersCoefficients}). The anharmonic correction has been estimated to -1.2~cm$^{-1}$ in Ref.~\cite{vanderbilt1986CalculationAnharmonicPhonon}.}
\vspace{5pt}
\centering
\begin{tabular}{M{2.0cm} M{2.0cm} M{2.0cm} M{2.0cm} M{2.1cm} M{2.1cm} M{1.0cm} M{1.6cm} M{1.8cm} M{1.4cm} M{2.8cm}}
    \toprule
     \makecell{\textbf{RPA} \\ \textbf{[OEP(33\%)]}} & \makecell{\textbf{RPAx} \\ \textbf{[OEP(33\%)]}} & \makecell{\textbf{RPAx(1)} \\ \textbf{[OEP(33\%)]}} & \makecell{\textbf{AC-SOSEX} \\ \textbf{[OEP(33\%)]}} & \makecell{\textbf{PBE0(33\%)} \\ \textbf{(scf)}} & \makecell{\textbf{PBE} \\ \textbf{(scf)}} &
     \makecell{\textbf{HF} \\ \textbf{(scf)}} &
     \makecell{\textbf{Expt.} \\ Ref.~\cite{nilsson1971PhononDispersionRelations}}
     \\
    \midrule
    305 & 309 & 306 & 310 & 310 & 295 & 336& 304\\
    \bottomrule
\end{tabular}
\end{table*}

In order to understand the impact of the choice of functional for determining the anharmonic shift, we have followed the same finite-difference approach as in Ref. \cite{vanderbilt1986CalculationAnharmonicPhonon} but with the RPAx-optimized hybrid functional, which we found to give similar results to RPAx for the phonon frequency. Using this method, we find the correction to be -14.2~cm$^{-1}$, only slightly smaller than that of PBE, -17.4~cm$^{-1}$. As such, the inclusion of exact exchange, crucial for the phonon frequency, appears to decrease the size of the anharmonic correction, and gives us a new experimental harmonic reference of 1348 cm$^{-1}$, which slightly decreases the difference with respect to RPA and slightly increases the difference with respect to RPAx.

Finally, the alternative RPAx resummations, RPAx(1) and AC-SOSEX, give us a spread of 20~cm$^{-1}$ of possible values when including the exact-exchange kernel. RPAx(1) predicts a harmonic frequency at 1348~cm$^{-1}$, which perfectly agrees with the experimental value, while AC-SOSEX predicts a value at 1368~cm$^{-1}$, which is 6~cm$^{-1}$ above the RPAx result. On the molecular systems studied in Sec.~\ref{secfiveA}, RPAx(1) systematically underestimated the vibrational frequencies, while AC-SOSEX overestimated them. RPAx also slightly overestimated the frequencies. 
If we assume that this trend holds for solids, the true frequency should lie somewhere between RPAx(1) and RPAx, i.e., in the interval 1348-1362~cm$^{-1}$. This implies an overestimation with respect to experiments of just a few~cm$^{-1}$. At the same time, it is reasonable to expect technical errors, such as the pseudopotential approximation, to be on the order of 5~cm$^{-1}$. 

Thanks to our force implementation in RPA, we have also calculated the full phonon dispersion. Since there is no impact of self-consistency, we have only considered RPA[PBE] to save computational cost.  We have used the same computational setup as before, only decreasing the number of eigenvalues per valence electron from 10 to 5, as the change in $\Gamma$-point frequency was found to be minimal, only affecting the frequency by 2 cm$^{-1}$ on the 6$\times$6$\times$6 grid. We then constructed a [222] diamond supercell containing 16 atoms. A supercell allows for a precise determination of the $\mathbf{q}$-points that are commensurate with its size, in our case, $\Gamma$, X, and L. The points between these high-symmetry $\mathbf{q}$-points are, however, approximated through a Fourier interpolation. For the choice of sampling, we used the same setting as our most advanced $\Gamma$-point calculation, which corresponds to a 3$\times$3$\times$3 $\mathbf{k}$-point grid on the supercell. The RPA phonon dispersion calculated on the experimental unit cell is given in Fig.~\ref{fig:Diamond_RPA_geom}, together with the results in PBE, PBE0(17\%), and PBE0(34\%). As can be seen, it is mostly the optical branches that are affected by the choice of functional.

The second material that we have considered in this study is the diamond phase of silicon. On this system, we have performed a similar study as for C. Previous works have shown that the lattice constant is rather sensitive to the pseudopotential approximation \cite{pitts2025SelfconsistentRandomPhase}. We have, therefore, first verified that the phonon frequencies do not show a similar sensitivity when evaluated at the experimental lattice constant. Using different pseudopotential libraries \cite{goedecker1996SeparableDualspaceGaussian,hartwigsen1998RelativisticSeparableDualspace}, we found discrepancies of maximum 2~cm$^{-1}$ for RPA and RPAx, and 3~cm$^{-1}$ for PBE. For this study, we have, therefore, continued to use the ONCV library.

We then computed the optical $\Gamma$-phonon using PBE, the RPA-optimized hybrid PBE0(29\%) \cite{pitts2025SelfconsistentRandomPhase}, RPA, RPAx, RPAx(1), and AC-SOSEX. For the PBE, PBE0(29\%), and RPAx calculations, we used a kinetic energy cutoff of 80~Ry. For RPA, 60~Ry was sufficient.
For the correlation energy, we used a grid of 8 imaginary frequency points and determined 60 eigenvalues of the response function in both RPA and RPAx. The total energy finite-difference calculations were obtained with mixed $\mathbf{k}$-point grids. 
The EXX energy was converged on a 8$\times$8$\times$8 $\G$-centered grid, while the correlation energy was converged on a shifted 4$\times$4$\times$4 grid. The value of the $\Gamma$ phonon frequency for all methods is listed in Tab.~\ref{tab:Tab_Silicon_Frequency}. The RPA[PBE] predicts a frequency at 522~cm$^{-1}$, in very good agreement with the experimental value at 523~cm$^{-1}$ ~\cite{Kulda1994}. Effects of self-consistency in RPA was again negligible. The RPAx frequency lies at 531~cm$^{-1}$, RPAx(1) at 527~cm$^{-1}$, and AC-SOSEX at 536~cm$^{-1}$. Using the same approach as before on C, we estimate the anharmonic shift of Si to be -3.1 cm$^{-1}$ with PBE and -2.9 cm$^{-1}$ with the optimized PBE0(29\%) hybrid, in good agreement with the value of -3.5~cm$^{-1}$ already presented in the literature \cite{vanderbilt1986CalculationAnharmonicPhonon}. Compared to the estimated harmonic experimental frequency at 526~cm$^{-1}$, the PBE underestimates the frequency by 12~cm$^{-1}$, the RPA underestimates by 3~cm$^{-1}$, RPAx/AC-SOSEX overestimate by 5/10~cm$^{-1}$, and RPAx(1) is exactly on top of the experimental value. This trend is the same as seen in diamond. 

The phonon dispersion is presented in Fig.~\ref{fig:Silicon_RPA_dispersion}. Similarly to diamond, it has been acquired using the forces on a [222] supercell. A 4$\times$4$\times$4 $\mathbf{k}$-point grid was considered for both PBE and PBE0(29\%). For RPA[PBE], we used a 3$\times$3$\times$3 sampling. At this sampling, the $\G$-mode agrees, within 2~cm$^{-1}$, with our fully converged total energy finite difference calculation presented above. Going from 2$\times$2$\times$2 to 3$\times$3$\times$3 resulted in very small changes in the dispersion plot, mainly around the L-point and for the first acoustic branch around the X-point. We are, therefore, confident that the phonon dispersion, in the vicinity of the high-symmetry $\mathbf{q}$-points, is converged in RPA[PBE] with a 3$\times$3$\times$3 grid on the supercell.

Finally, we have investigated the diamond phase of germanium. This material has a small indirect gap of 0.66 eV \cite{doi:10.1021/jacs.8b03503}. At the optimized lattice parameter, PBE predicts a band overlap, but at the experimental lattice parameter (5.6575~\!\AA~\cite{straumanis1952LatticeParametersCoefficients}), a small gap opens ($<$0.1 eV). Including a fraction of exact-exchange opens the gap further. Due to the change in nature of the ground state, we expect larger effects of self-consistency in Ge. Indeed, evaluating EXX on top of PBE orbitals leads to a slow and irregular convergence. This behavior is strongly improved by evaluating EXX on top of a hybrid PBE0 functional. In order to find a starting point that gives a density close to self-consistent RPA, we optimized the PBE0$\a$ functional using RPA, resulting in 33\% of exact exchange. The corresponding OEP(33\%) method, abbreviation of OEP-PBE0(33\%), yields a KS gap of 0.32 eV, which agrees well with the self-consistent RPA gap of 0.28 eV.

The $\G$-phonon was then evaluated at the experimental lattice parameter, with the total energy finite-difference approach, for RPA[OEP(33\%)] and self-consistent RPA. At a $\G$-centered 4$\times$4$\times$4 $\mathbf{k}$-point-grid, we found a difference of only 2~cm$^{-1}$ in the phonon frequency between the two methods. We, therefore, continued using  OEP(33\%) as a starting point for RPAx, RPAx(1), and AC-SOSEX. We found a quick convergence with respect to the number of eigenvalues of the response function. Results were well-converged already with a total of 20-40 eigenvalues. On the other hand, due to the smaller gap in Ge, we needed to include up to 12 imaginary frequency points. A plane-wave cutoff of 60 Ry was sufficient. For the $\mathbf{k}$-point convergence, we used a similar setup as in silicon: 8$\times$8$\times$8 for the EXX part of the energy and a 4$\times$4$\times$4 shifted grid for the correlation energy. The results for the $\G$-phonon are presented in Tab. \ref{tab:Tab_Germanium_Frequency}. We have also included the frequency in PBE0(33\%) and PBE, where we used a 8$\times$8$\times$8 $\mathbf{k}$-point grid and a 80~Ry plane-wave cutoff. 

The anharmonic correction was estimated also for Ge in Ref.~\cite{vanderbilt1986CalculationAnharmonicPhonon}, giving a value of -1.2~cm$^{-1}$. Taking this into account, the experimental harmonic phonon is at 305~cm$^{-1}$. The RPA[OEP(33\%)] value is exactly on top of this value. We see that RPA corrects the PBE frequency by 10~cm$^{-1}$. The effect of the exact-exchange kernel is rather small, increasing the RPA frequency by 4~cm$^{-1}$ in RPAx, 1~cm$^{-1}$ in RPAx(1), and 5~cm$^{-1}$ in AC-SOSEX.  Thus, every ACFDT-based method predicts a value in good agreement with experiments. The PBE0(33\%) functional finds the optical phonon at 310~cm$^{-1}$, which, similar to C and Si, slightly overestimates the RPA result.

We have not included the Ge dispersion plot in RPA due to the slow $\mathbf{k}$-point convergence of the phonons around the L-point, which would necessitate larger grids than 3$\times$3$\times$3 on the [222] supercell. In addition, a self-consistent calculation would be needed, which multiplies the cost by the number of iterations required to converge the KS orbitals. 

\section{Conclusions and outlook}

In this work, we have derived and implemented atomic forces within the RPA method. Forces were evaluated at full self-consistency using the OEP method, and non-self-consistently using density functional perturbation theory with the PBE functional. Both self-consistent and non-self-consistent forces were shown to be of high numerical quality. Similarly to EXX and hybrid functionals, a crucial extra force term remains at full self-consistency when nonlocal pseudopotentials are employed. 

The RPA force implementation has been applied to molecular and solid-state systems, highlighting a systematic improvement of structural and vibrational properties with respect to simpler semi-local functionals. The errors in RPA, in particular for molecules, are found mainly in bond angles and bond-stretching modes, which are not as well described as expected. On the other hand, in diamond, RPA strongly 
improves the zone-center optical phonon. The error, taking into account the anharmonic shift, drops from 4.3\% in PBE to 0.5\% in RPA. Similar improvements are found for silicon and germanium. Overall, the error on the vibrational frequencies in RPA seems smaller in solids than in molecules.

Despite the current analytical force implementation being available only for RPA, the finite-difference total energy estimates made at the RPAx level show an improvement over RPA. In fact, an excellent agreement with CCSD(T) is found for all molecules. On diamond, our RPAx phonon calculation provides the best theoretical estimate obtained so far. Future works should therefore aim to implement the exact-exchange kernel contribution to the gradient, to gain access to RPAx analytical forces. 
A non-self-consistent implementation should be sufficient, as demonstrated for RPA in this work.

Finally, the current RPA force implementation represents an important step towards the analytical calculation of second-order derivatives of the total energy with respect to atomic positions, and electron-phonon couplings. Another important future direction is the evaluation of the stress tensor, which would enable full cell relaxation.
\newpage
\section*{Appendix}
\begin{figure}[t]
\begin{center}
\includegraphics[scale=0.40,angle=0]{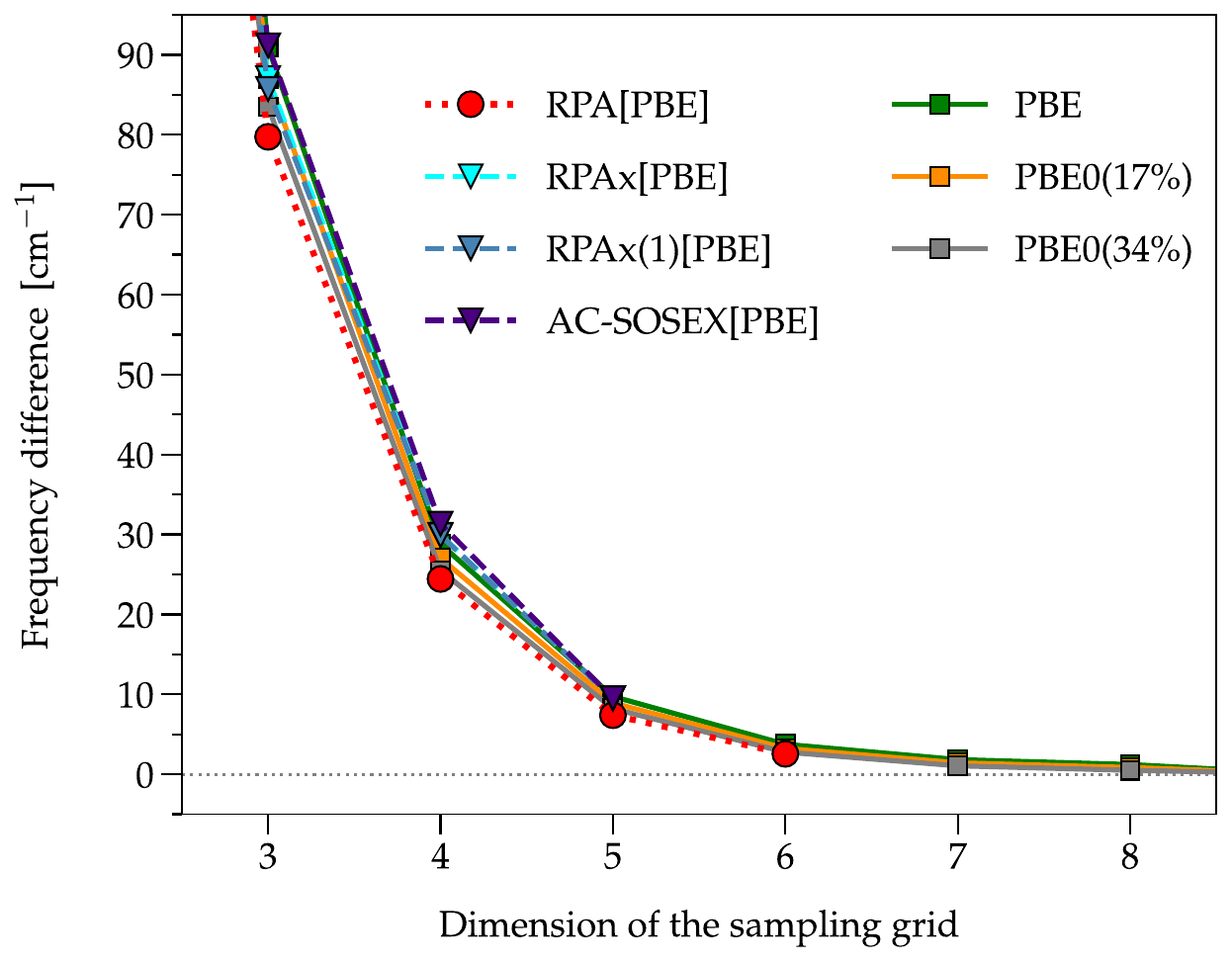}
\end{center}
\caption{\label{fig:Convergence_kpoints_vBIS} Convergence of the $\Gamma$-point optical phonon frequency of diamond with respect to the size of the sampling grid. The difference in frequency evaluated at a given $\mathbf{k}$-point and $\mathbf{q}$-point grid definition is given with respect to the frequency calculated on a 5$\times$5$\times$5 grid for RPAx, RPAx(1), AC-SOSEX, a 6$\times$6$\times$6 grid for RPA, and a 10$\times$10$\times$10 grid for PBE, PBE0(17\%), and PBE0(34\%). The RPA-based methods' curves have been shifted upwards for extrapolation. The ACFDT calculations were performed with a kinetic
energy cutoff of 80 Ry, and a total of 80 eigenvalues.}
\end{figure}

\begin{figure}[t]
\begin{center}
\includegraphics[scale=0.40,angle=0]{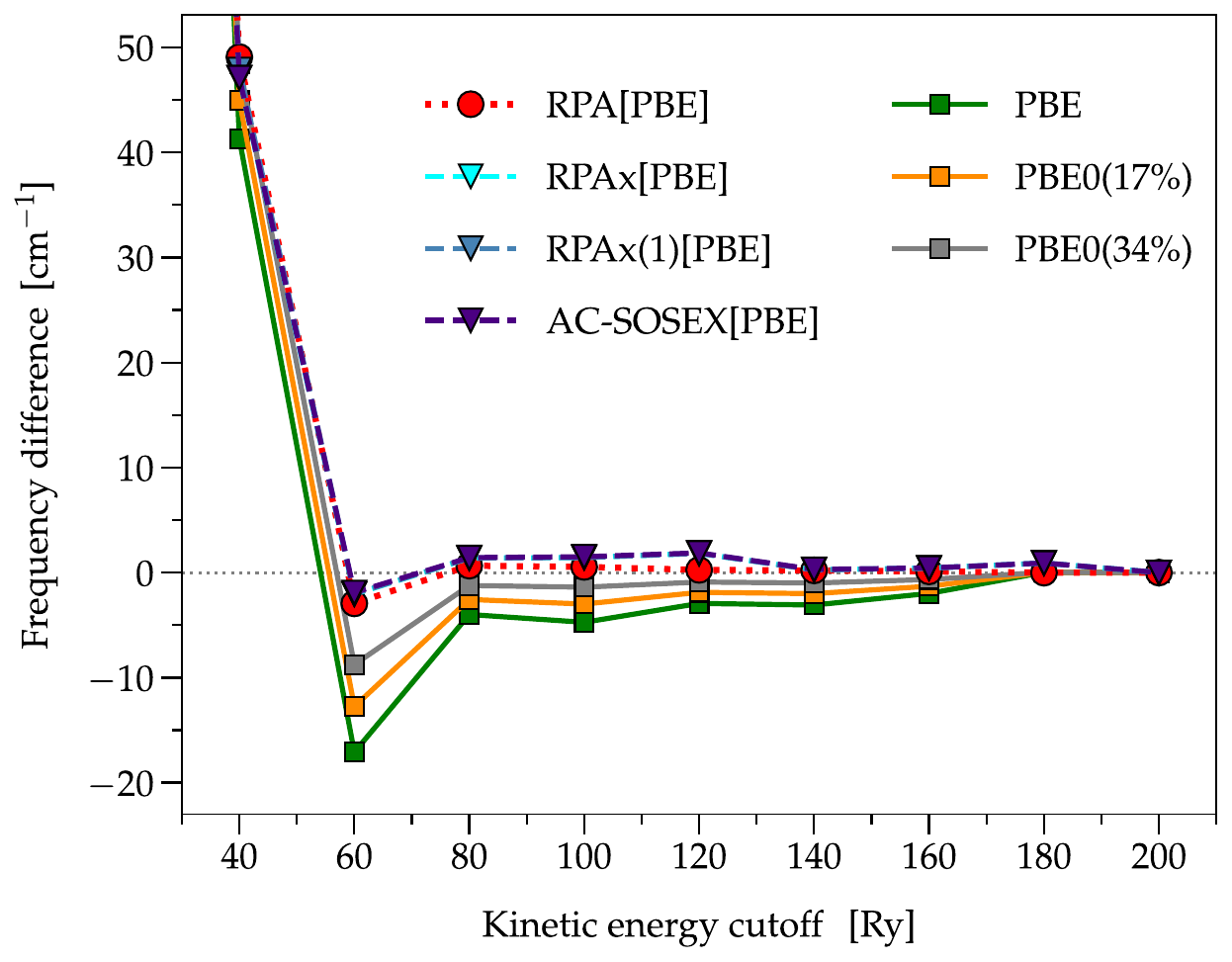}
\end{center}
\caption{\label{fig:Convergence_cutoff_vBIS} Convergence of the $\Gamma$-point optical phonon frequency of diamond with respect to the kinetic energy cutoff. The difference in frequency evaluated for a given cutoff is given with respect to the value calculated at 200~Ry. The PBE, PBE0(17\%), and PBE0(34\%) calculations have been performed on a 8$\times$8$\times$8 $\mathbf{k}$-point grid, while the ACFDT-based methods have been performed on a 3$\times$3$\times$3 grid. A total number of 80 eigenvalues was used.}
\end{figure}

In this Appendix, we present convergence studies for diamond. The $\mathbf{k}$-point convergence is presented in Fig.~\ref{fig:Convergence_kpoints_vBIS}. The behavior is very similar for all methods studied, thus allowing us to extrapolate the RPA, RPAx, RPAx(1), and AC-SOSEX results with high precision. 
The convergence with respect to the plane-wave cutoff is presented in Fig.~\ref{fig:Convergence_cutoff_vBIS}. The ACFDT-based methods are seen to converge faster with respect to this parameter, as compared to PBE and hybrid functionals. 

\begin{figure}[h]
\begin{center}
\includegraphics[scale=0.40,angle=0]{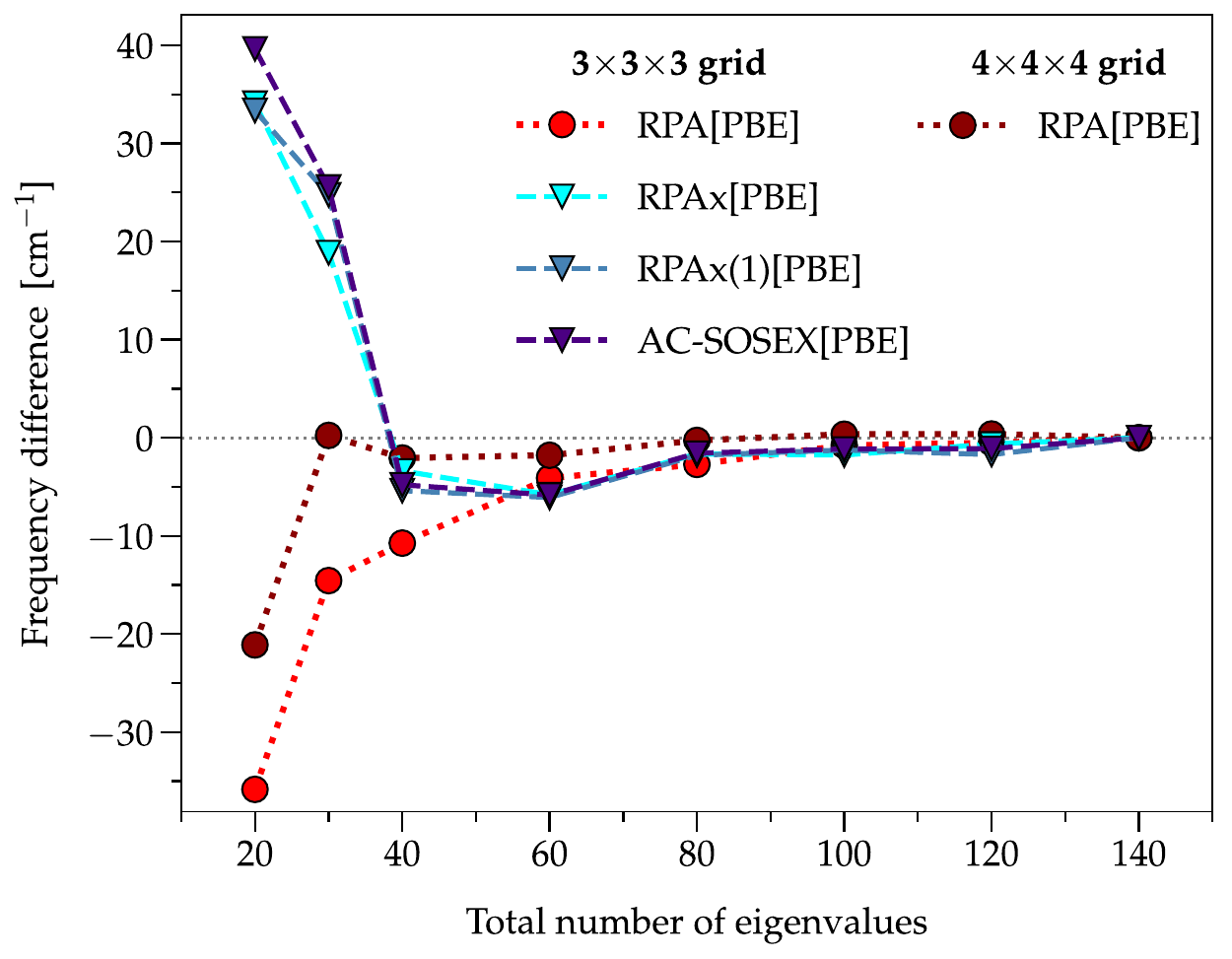}
\end{center}
\caption{\label{fig:Convergence_eigenvalues_vBIS} Convergence of the $\Gamma$-point optical phonon frequency of diamond with respect to the total number of eigenvalues of the response function. The difference in frequency evaluated for a given number of eigenvalues is taken with respect to the frequency calculated at 140 eigenvalues. Calculations have been performed on a 3$\times$3$\times$3 and a 4$\times$4$\times$4 $\mathbf{k}$-point grid. A plane-wave cutoff of 80~Ry was used.}
\end{figure}

The ACFDT correlation energy requires a convergence with respect to the number of eigenvalues generated for the response function, $\chi_0$. In Fig.~\ref{fig:Convergence_eigenvalues_vBIS}, we see that this convergence is rapid for all approximations. We also see that it improves in RPA using larger $\mathbf{k}$-point grids.

\begin{figure}[h]
\begin{center}
\includegraphics[scale=0.40,angle=0]{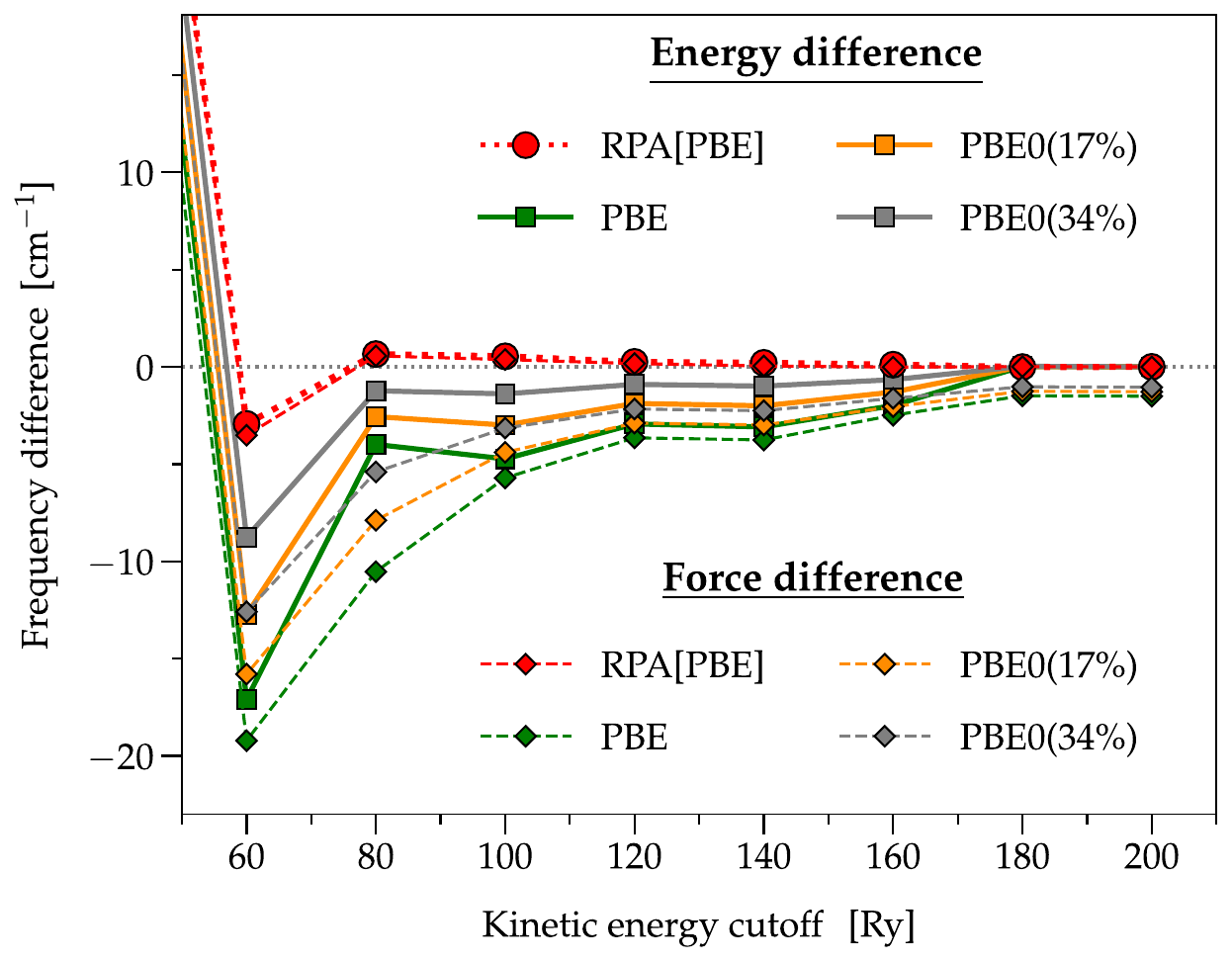}
\end{center}
\caption{\label{fig:Difference_ecutoff_FD_phonopy} Difference in the $\Gamma$-point optical phonon frequency estimated by finite difference of the total energy and by finite difference of forces via the Phonopy software. The calculations have all been performed considering a 3$\times$3$\times$3 $\mathbf{k}$-point grid, and a total number of 80~eigenvalues for RPA[PBE]. The curves have all been shifted with respect to the RPA[PBE] frequency value calculated by finite difference at 200~Ry.}
\end{figure}

In Fig.~\ref{fig:Difference_ecutoff_FD_phonopy}, we see that for the smaller values of the cutoff, a discrepancy exists between the total energy finite-difference and force finite-difference estimates. This behavior disappears, however, when increasing the value of the cutoff. While RPA[PBE] is barely impacted, the PBE and PBE0 hybrid functional are. For them, a perfect agreement is only reached at very large cutoff values.

\acknowledgements
The work was performed using HPC resources from GENCI-TGCC/CINES/IDRIS (Grant No. A0150914650). 
Discussions with Dr. Michele Casula are acknowledged.
\providecommand{\noopsort}[1]{}\providecommand{\singleletter}[1]{#1}
\end{document}